\documentclass[10pt,twocolumn,tightenlines,aps,amsmath,floatfix,notitlepage,nofootinbib,showpacs,prb,superscriptaddress]{revtex4-1}

\usepackage[T1]{fontenc}        
\usepackage{palatino}           
\usepackage{amsmath}            
\usepackage{amssymb}            
\usepackage{graphicx}           
\usepackage{rotating}
\usepackage{booktabs}
\usepackage{hyperref}           
\usepackage{pslatex}
\usepackage{epsfig}
\usepackage{epsf}
\usepackage[usenames]{color}
\usepackage{mathrsfs}           
\usepackage{mathtools}          
\newcommand{\sfrac}[2]{\frac{#1}{#2}}
\newcommand{\V}[1]{\mathbf{#1}}
\newcommand{\bra}[1]{\langle #1|}
\newcommand{\ket}[1]{|#1\rangle}

\newcommand{\expbk}[1]{\langle #1 \rangle}
\newcommand{\rem}[1]{}          
\def\app#1#2{%
  \mathrel{%
    \setbox0=\hbox{$#1\sim$}%
    \setbox2=\hbox{%
      \rlap{\hbox{$#1\propto$}}%
      \lower1.1\ht0\box0%
    }%
    \raise0.25\ht2\box2%
  }%
}
\def\appropto{\mathpalette\app\relax}

\renewcommand{\ij}{\mathrlap{i}j}
\newcommand{\clJ}{\mathcal{J}}
\newcommand{\clK}{\mathcal{K}}
\newcommand{\nn}{\mathrm{nn}}
\newcommand{\nnn}{\mathrm{nnn}}
\newcommand{\nnc}{\mathrm{nnc}}
\newcommand{\ntc}{\mathrm{n3c}}
\newcommand{\tmf}{$\times 10^{-4}$}
\newcommand{\tmv}{$\times 10^{-5}$}

\begin{document}

\title{A Mean Field Model for the Quadrupolar Phases of UPd$_3$}

\author{Manh Duc Le}
\affiliation{Center for Correlated Electron Systems, Institute for Basic Science (IBS), Seoul 151-747, Korea}
\affiliation{Dept. of Physics and Astronomy, Seoul National University, Seoul 151-747, Korea}
\affiliation{Dept. of Physics and Astronomy, and London Centre for Nanotechnology, University College London, Gower Street, London, WC1E 6BT, UK}
\author{Keith A. McEwen}
\affiliation{Dept. of Physics and Astronomy, and London Centre for Nanotechnology, University College London, Gower Street, London, WC1E 6BT, UK}
\author{Martin Rotter}
\affiliation{Max Planck Institute for Chemical Physics of Solids, N\"othnitzer Str. 40, D-01187 Dresden, Germany}
\affiliation{Institut f\"ur Physikalische Chemie, Universit\"at Wien, W\"ahringerstr. 42, 1090 Wien, Austria}
\author{Mathias Doerr}
\affiliation{Institut f\"ur Festk\"orperphysik, Technische Universit\"at Dresden, D-01062 Dresden, Germany}
\author{Alexander Barcza}
\affiliation{Institut f\"ur Physikalische Chemie, Universit\"at Wien, W\"ahringerstr. 42, 1090 Wien, Austria}
\affiliation{Vakuumschmelze GmbH, Gr\"unerweg 37, D-63450 Hanau, Germany}
\author{Je-Geun Park}
\affiliation{Center for Correlated Electron Systems, Institute for Basic Science (IBS), Seoul 151-747, Korea}
\affiliation{Dept. of Physics and Astronomy, Seoul National University, Seoul 151-747, Korea}
\author{James Brooks}
\affiliation{National High Magnetic Field Laboratory, 1800 Paul Dirac Dr., Tallahassee, Florida, 32306, USA}
\author{Eric Jobiliong}
\affiliation{National High Magnetic Field Laboratory, 1800 Paul Dirac Dr., Tallahassee, Florida, 32306, USA}
\affiliation{Department of Industrial Engineering, Universitas Pelita Harapan Karawaci, Banten, 15811, Indonesia}
\author{David Fort}
\affiliation{Department of Metallurgy and Materials, University of Birmingham, Birmingham, B15 2TT, UK}

\date{\today}

\begin{abstract}

UPd$_3$ is known to exhibit four antiferroquadrupolar ordered phases at low temperatures. We report measurements of the magnetisation and
magnetostriction of single crystal UPd$_3$, along the principal symmetry directions, in fields up to 33~T. These results have been combined
with recent inelastic neutron and x-ray resonant scattering measurements to construct a mean field model of UPd$_3$ including up to fourth
nearest neighbour interactions. In particular we find that anisotropic quadrupolar interactions must be included in order to explain the low
temperature structures derived from the scattering data.

\end{abstract}

\pacs{75.80.+q, 75.40.Cx, 75.10.Dg, 71.10.Hf} 

\maketitle

\section{Introduction} \label{sec-upd3-intro}

Low temperature phase transitions in condensed matter systems are usually driven by the cooperative actions of the system's electronic degrees of
freedom. In many cases the exchange interactions between the spins of electrons on neighbouring ions causes them to order below a characteristic
transition temperature. However, in lanthanide compounds, the localised $4f$-electrons on each ionic site may have non-spherical charge or current
distributions, which may be described by a multipole expansion of their electric or magnetic fields~\cite{rmp_multipoles}. These multipoles may
interact, and in certain cases these interactions may be stronger than the spin exchange interactions, driving a phase transition to a multipolar
ordered phase. There have been many examples of electric quadurpolar order observed, such as in CeB$_6$~\cite{ceb6xrayEU,ceb6xrayjap},
PrPb$_3$~\cite{onimaru_prpb3}, and TmTe~\cite{nagao_tmte}, whilst higher order electric multipoles were observed in DyB$_2$C$_2$~\cite{tanaka_dyb2c2}.

\begin{figure*}
  \begin{center}
    \includegraphics[width=\textwidth,viewport=20 17 482 203]{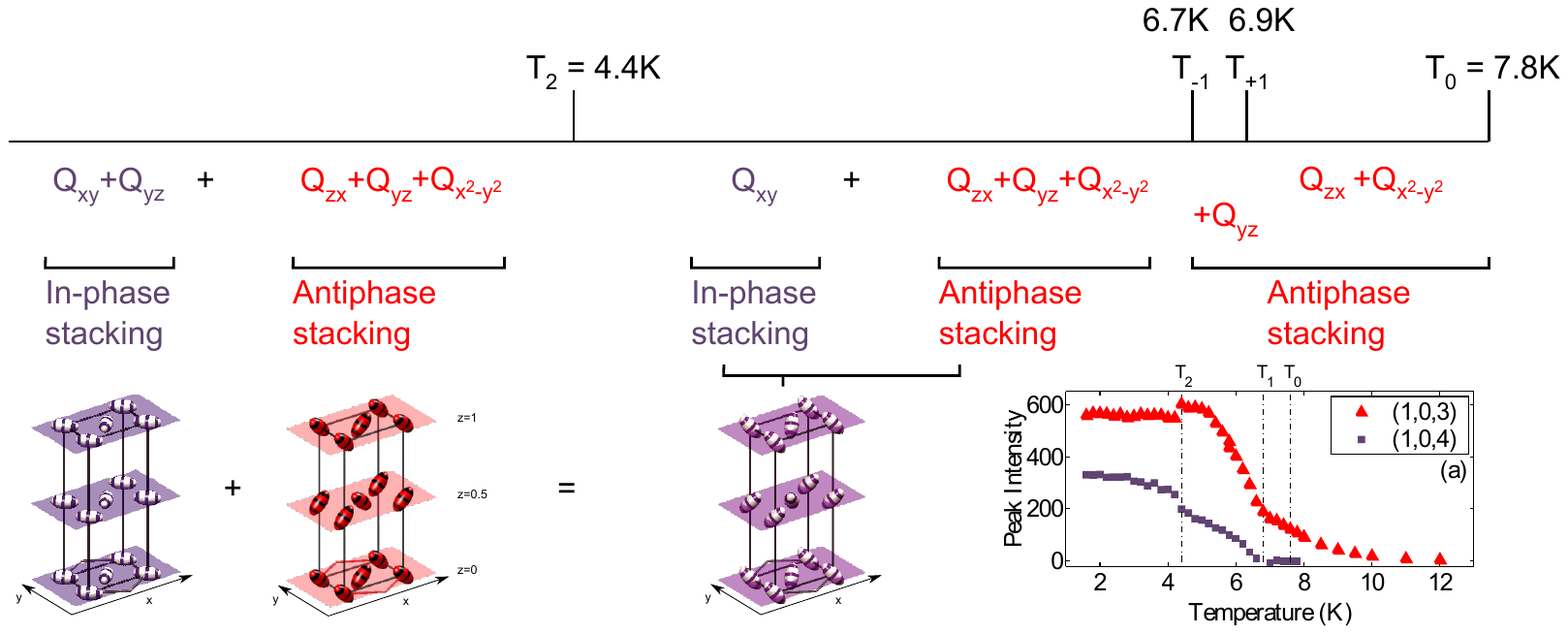}
    \caption{(Color Online) \emph{The quadrupolar phases of UPd$_3$}. Labels below the line indicate order parameter as determined from XRS data. `In phase'
	     indicates the quadrupole moments are aligned in parallel along the c-axis. This corresponds to a superlattice reflection at Q=(1~0~4).
             `Anti-phase' indicates an anti-parallel alignment along c, and corresponds to a reflection at Q=(1~0~3). After Walker et al.\cite{WMM+06,hcwESRF07}}
    \label{fg:ordertemps} \end{center}
\end{figure*}

In contrast to the lanthanides, there have not been many cases of multipolar order observed in actinide compounds. This is due partly to the
larger spatial extent of the $5f$ wavefunctions, which causes them to become delocalised, and hence invalidates any description of their
electric or magnetic fields about a particular point. The insulating actinide oxides UO$_2$ and NpO$_2$, however, have been studied
extensively, and exhibit electric quadrupolar and high order magnetic multipolar order respectively~\cite{rmp_multipoles}. In contrast
UPd$_3$ is one of the few metallic actinide compounds which has well localised $5f$ electrons, and it was one of the first compounds to be
shown to exhibit quadrupolar ordering. Anomalies were observed in the heat capacity~\cite{ZdPM95}, thermal expansion~\cite{ZdPM95},
magnetic susceptibility~\cite{MPGG03} and electrical resistivity~\cite{tokiwa01} of UPd$_3$ at low temperatures which are indicative of
phase transitions. Polarised neutron diffraction measurements revealed superlattice peaks at $\V{Q}=(\frac{1}{2} 0 l)$ which appear only in
the non-spin-flip channel, and thus can be attributed to a structural transition~\cite{SMMF92}, from hexagonal to orthorhombic symmetry.

X-ray resonant scattering (XRS) measurements~\cite{MMS+01} showed that this transition arises from the ordering of the electric quadrupole
moment of the 5$f^2$ electrons. In addition to this phase transition at $T_0=7.8$~K, there are three further transitions to different
antiferroquadrupolar (AFQ) ordered states at $T_{+1}=6.9$~K, $T_{-1}=6.7$~K and $T_2=4.4$~K. Below $T_0$ a superlattice peak at
$(\frac{1}{2} 0 l)$ where $l$ is odd is observed, whilst below $T_{-1}$ there are additional peaks at $(\frac{1}{2} 0 l)$ where $l$ is even.
The $l$ odd peaks show that there is antiferroquadrupolar ordering along the $c$-direction, also denoted as an anti-phase stacking of
quadrupoles. The $l$ even peaks show an additional ordering of quadrupole moments in-phase along $c$.

Measurements of the order parameter using x-ray resonant scattering show that the $l$ odd order is associated mainly with $Q_{zx}$
quadrupoles~\cite{WMM+06}, whilst the $l$ even order is associated with $Q_{xy}$ quadrupoles~\cite{hcwESRF07}. The $Q_{zx}$ ordering is
accompanied by a component from the $Q_{x^2-y^2}$ quadrupoles, whilst the $Q_{xy}$ is accompanied by a $Q_{yz}$ component. The directions
$x$ and $z$ are equivalent to the $a$ and $c$ crystallographic directions and $y$ is perpendicular to both. In addition, there is also an
additional ordering of the $Q_{yz}$ quadrupoles in anti-phase below $T_{+1}$. This sequence of phase transitions is summarised in
figure~\ref{fg:ordertemps}.

UPd$_3$ adopts the double hexagonal close packed TiNi$_3$ crystal structure (space group $D_{6h}^4$, $P6_3/mmc$, no. 194)~\cite{HW55} with
lattice parameters $a=5.76$~\AA~and $c=9.62$~\AA. This means that the nearest neighbour U-U distance, 4.1~\AA, is larger than the Hill limit
($\approx 3.5$~\AA)~\cite{Hill70} and thus the 5$f^2$ electrons are well localised. The U$^{4+}$ ions are in the $2a$ and $2d$ sites, which
have respectively $D_{3d}$ ($\bar{3}m$) and $D_{3h}$ ($\bar{6}m2$) point symmetry which we shall refer to as \emph{quasi-cubic} and
\emph{hexagonal}. The XRS data shows that the ordering involves primarily the ions on the quasi-cubic sites~\cite{WMM+06,hcwESRF07}. 

Knowledge of the crystal field (CF) interactions is essential in determining the more complex two-ion interactions. In particular it is
crucial to know the CF ground state. Moreover, the excited states and the matrix elements of the angular momentum operators, $\hat{J}_i$,
between these and the ground state determine the intensities of excitations observed by inelastic neutron scattering and also to some extent
the magnitude of the magnetisation in the ordered phase, as explained by McEwen et al.~\cite{MPGG03}. Thus we shall first consider in
section~\ref{sec-upd3-cf} the CF level scheme deduced from inelastic neutron scattering measurements in the paramagnetic phase.

\begin{figure}
\begin{center}
  \includegraphics[width=0.9\columnwidth,viewport=0 0 297 339]{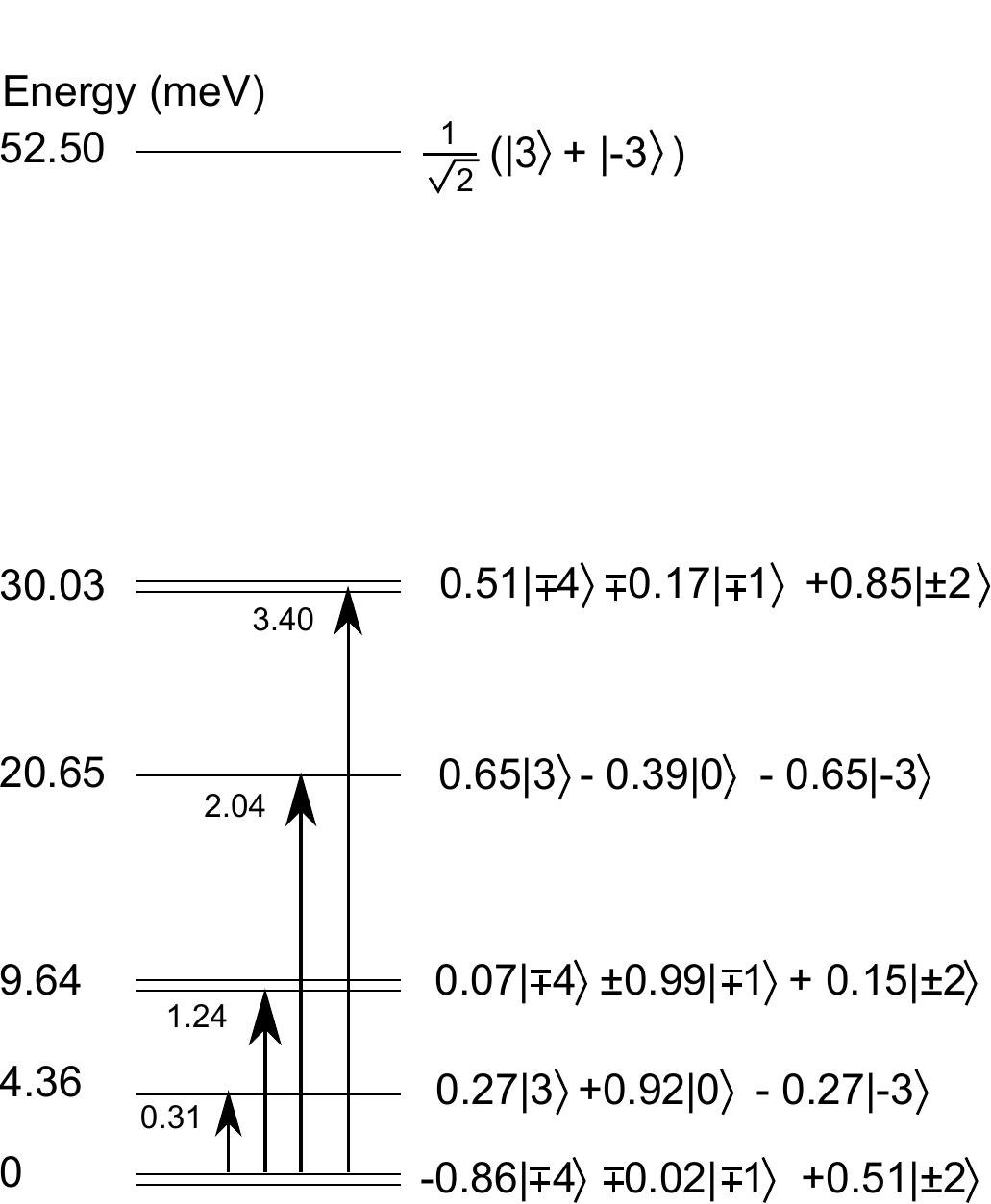}
  \caption{\emph{U$^{4+}$ quasi-cubic site crystal field Level Scheme.} The crystal field energy levels and corresponding wavefunctions in 
           the paramagnetic phase expressed in the $\ket{J=4,J_z}$ basis. Arrows denote transitions from the ground state with non-negligible 
           dipole matrix elements whose squared values are shown as numbers near the arrow.} 
  \label{tab:CFscheme} 
\end{center}
\end{figure}

We then present high field magnetisation and magnetostriction measurements in section~\ref{sec-highfield}, from which the critical fields
and magnetic phase diagrams were determined experimentally. Finally, a mean field model of the quadrupolar ordering is constructed in
section~\ref{sec-upd3-mfcub}. The order parameters of each of the quadrupolar phases (determined by resonant X-ray scattering) and their
transition temperatures, were used to constrain the two-ion quadrupolar interaction parameters for the model, whilst the dipolar exchange
parameters were determined from the measured critical fields. The results of the model are then compared to the measured high field data.

Whilst the model which we shall present is empirically based on the measured physical properties of UPd$_3$ rather than directly on measured
exchange constants, we note that it is the first attempt to explain comprehensively this fascinating compound with its many competing
ordered phases from a microscopic point of view, and hope to stimulate further \emph{ab initio} studies of the exchange interactions
involved. The quadrupolar interactions which drive the many phase transitions in UPd$_3$ are difficult to measure directly, because neutrons
couple only to the magnetic dipoles in the system, and the energies are far too low for inelastic X-ray scattering to resolve. However, the
wealth of physical property measurements available on UPd$_3$ has encouraged us to try to synthesise this into a mean-field model which
explains, to a large extent, these varied measurements. We hope that this may encourage the construction of models to explain the properties
of similar quadrupolar (or higher multipolar) ordered compounds thus deepening the understanding of what lies behind these phenomena.

\section{Crystal Field Interactions} \label{sec-upd3-cf}

The single-ion properties of uranium are generally found to be close to the $LS$-coupling limit~\cite{moorevanderlaan}. So, in order to simplify the
analysis, we shall ignore any mixing with higher order multiplets in determining the crystal field (CF) parameters. As mentioned in
section~\ref{sec-upd3-intro}, there are two inequivalent sites for the U$^{4+}$ ions in the crystal structure of UPd$_3$. The different
point symmetry of these sites gives rise to different crystal fields, but it happens that both split the 9-fold degenerate $J=4$ ground
multiplet into three singlets and three doublets. The energies and wavefunctions of these levels, however, are different for the two sites.

The nature of the ground state may be deduced from single crystal susceptibility measurements and estimates of the magnetic entropy
determined from heat capacity measurements. These results indicate a singlet ground state on the hexagonal sites and a doublet on the quasi-cubic
sites~\cite{MPGG03}.

The CF split energy levels were determined from previously reported inelastic neutron scattering measurements made on the time-of-flight spectrometer 
HET at the ISIS Facility, UK~\cite{jmmm_upd3}, and on the triple-axis-spectrometer IN8 at ILL, Grenoble~\cite{amm00}. We identified magnetic excitations at 4.1, 9.7, 12.3,
16.8, 20.4 and 30~meV. The 16~meV peak exhibits considerable dispersion and is assigned to the transition between the $\ket{J_z=0}$ singlet ground
state and the $\ket{J_z=\pm1}$ doublet excited state of the hexagonal site ions. Its dispersion was used to determine the exchange
interactions between hexagonal sites~\cite{upd3ins}. The remaining peaks are assigned to the quasi-cubic site ions. 

Henceforth we shall be concerned mainly with the quasi-cubic sites, as the quadrupolar ordering primarily involves the uranium ions on these sites. 
These sites have trigonal, $\bar{3}m$ ($D_{3d}$), point symmetry, so that the crystal field Hamiltonian is

\begin{equation} \label{eq:Hcfcub}
\mathcal{H}_{\mathrm{cf}} = \sum_{k=2,4,6} B_k^0 O_k^0 + \sum_{k=4,6} B_k^3 O_k^3 + B_6^6 O_6^6
\end{equation}

\noindent where $B_k^q$ are crystal field parameters and $O_k^q$ are Stevens operators. The quantisation ($z$) axis is taken to be the trigonal axis,
which in this case is parallel to $c$. 

From the measured transition energies and with the restriction of a doublet ground state, a crystal field fitting program~\cite{NN00} was
used to obtain initial estimates of the CF parameters for the quasi-cubic sites. This program relies on the orthogonality of the spherical harmonic
functions from which the CF operators are constructed. It allows one to find a set of CF parameters, $B_k^q$, given the energy levels and wavefunctions
produced by the crystal field. The fitting algorithm may thus vary either the wavefunctions to fit a particular set of energy levels, or vice versa. 
In this case, however, we also face constraints on the wavefunctions. 

\begin{table} \renewcommand{\arraystretch}{1.3}
\begin{center}
  \begin{tabular}{@{\extracolsep{\fill}}rl|rl}
   \hline 
   $B_0^2$    &  0.035 ~   & ~ $B_0^6$ & -0.00012  \rem{|& B$_0^2$ &  0.21     & B$_0^6$ &  0.00033  |}\\
   $B_0^4$    & -0.012 ~   & ~ $B_3^6$ &  0.0025   \rem{|& B$_0^4$ & -0.0117   & B$_6^6$ &  0.0014   |}\\ 
   $B_3^4$    & -0.027 ~   & ~ $B_6^6$ &  0.0068   \rem{|&         &           &         &           |}\\ \hline 
  \end{tabular}
  \caption{\emph{Crystal field parameters in Stevens normalisation in meV}. The parameters were deduced from fitting to inelastic neutron spectra and
           the constraints on the wavefunctions of the lowest three energy levels of the quasi-cubic sites as described in the text.} 
  \label{tab:CFpars}
\end{center}
\end{table}

\begin{figure}
  \begin{center}
    \includegraphics[width=0.95\columnwidth,viewport=73 185 493 612]{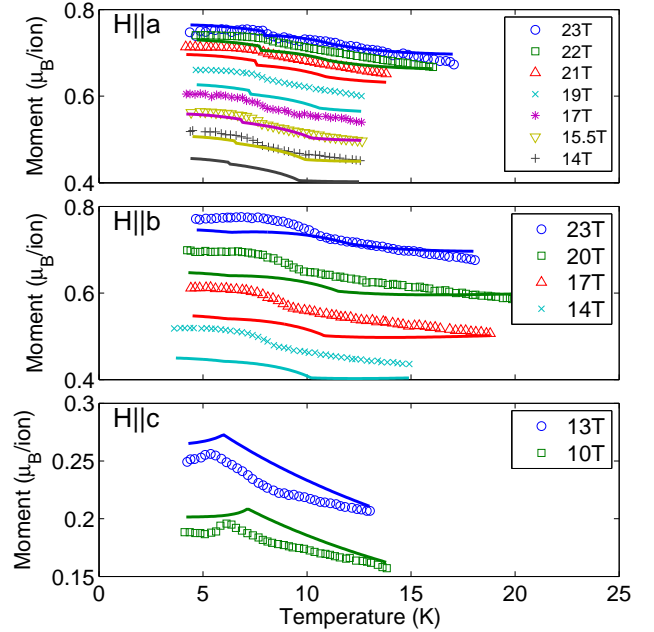}
    \caption{(Color Online) \emph{The high field magnetisation of UPd$_3$ as a function of temperature.} Solid lines are calculated from the mean-field
	     model. The calculated values in the bottom panel (for field parallel to $c$) have been divided by 3.} 
    \label{fg:highfield-mag}
  \end{center} 
\end{figure}

\begin{figure}[t]
  \begin{center}
    \includegraphics[width=1.00\columnwidth,viewport=90 62 504 711]{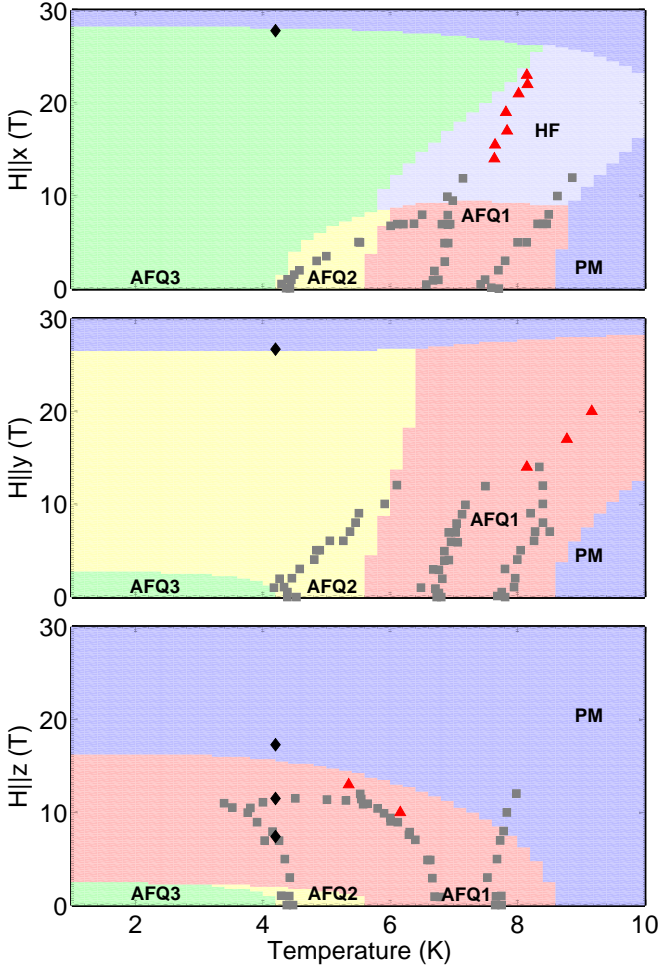}
    \caption{(Color Online) \emph{Magnetic Phase diagrams of UPd$_3$ for fields parallel to $a$, $b$ and $c$.} Filled squares are from previous 
              measurements by McEwen et al.~\cite{MU+98} Filled triangles are from high field magnetisation measurements described in this work, whilst 
              filled diamonds are from high field magnetostriction. The shaded background shows the \emph{McPhase} calculated phase diagrams, with zero 
              field phases AFQ1-AFQ3 as described in the text. PM is paramagnetic (or ferromagnetically polarised high field) phase, and HF is the high 
              field phase for $H||a$ which has the ordering wavevector $(\frac{1}{2}\frac{1}{2}0)$ with all quadrupolar moments ordered.}
    \label{fg:phasediagram}
  \end{center}
\end{figure}

From symmetry considerations, the doublet ground states have the wavefunctions

\begin{eqnarray} \label{eq:doubletGS}
\ket{d_1} &=& a\ket{4} + b\ket{1} + c\ket{-2} \\ \nonumber
\ket{d_2} &=& a\ket{-4} - b\ket{-1} + c\ket{2}
\end{eqnarray}

\noindent where for brevity the kets denote states $\ket{J=4,J_z}$. The singlet wavefunctions have the forms

\begin{eqnarray} \label{eq:singletES}
\ket{s} = d\ket{3} + e\ket{0} - d\ket{-3} \\ \nonumber
\ket{s'}=\frac{1}{\sqrt{2}} \left(\ket{3} + \ket{-3}\right)
\end{eqnarray}

In order to ensure that the T$_0=7.8$~K transition is accompanied by only a very small entropy change, as deduced from the heat capacity
data, the Landau theory analysis of McEwen et al.\cite{MPGG03} requires that the matrix elements $\bra{d_1} Q_{zx}\ket{d_2} = \bra{d_2}
Q_{zx} \ket{d_1} \approx 0$, where $Q_{zx}=\frac{1}{2}(J_xJ_z+J_zJ_x)$. As shown in the reference, this implies that $bc \approx 0$.

In addition, we note that the basal plane susceptibilities $\chi_{x,y}$ increase with decreasing temperature through the $T_{-1}=6.7$~K and
$T_2=4.4$~K transitions. This may be explained if the first excited state is a singlet and there is a large $\hat{J}_{x,y}$ matrix element
between it and some higher energy state which increases the $x-$ or $y-$direction susceptibility in the ordered phases as progressively more
of the singlet state is mixed in with the doublet ground state~\cite{gilliannote}. As the $\hat{J}_{x,y}$ matrix elements between singlet
states are zero, this coupling must be to a higher lying doublet, $\ket{d_{1,2}^{(2)}}$. The condition that
$\bra{s}\hat{J}_{x,y}\ket{d_{1,2}^{(2)}}$ be large whilst $\bra{s}\hat{J}_{x,y}\ket{d_{1,2}}$ is small is thus satisfied if $e\approx1$,
$b^{(2)}\approx1$ and $b\approx 0$.

These requirements are satisfied by the crystal field parameters in table~\ref{tab:CFpars}, which yield $b=0.02$, $b^{(2)}=0.99$ and
$e=0.92$. The parameters were obtained using a simulated annealing minimisation procedure whereby at each iteration, the algorithm mentioned
above~\cite{NN00} was used to refine an initial set of parameters to fit the measured energy levels. These refined parameters are
subsequently used to calculate the $b$, $b^{(2)}$ and $e$ matrix elements, from which the simulated annealing `energy' is obtained, and
hence minimised. Figure~\ref{tab:CFscheme} shows the resulting crystal field energy splitting and wave functions for the U$^{4+}$ ions on
the quasi-cubic sites.

\section{High Magnetic Field Measurements} \label{sec-highfield}

Single crystals were grown by the Czochralski technique at the University of Birmingham, and cut with faces perpendicular to the orthogonal
axes $x$, $y$ and $z$, where $x\|a$ and $z\|c$. These were used in magnetisation and magnetostriction measurements in fields up to 14~T at
Birkbeck College~\cite{MU+98} by two of us (KAM and JGP). Subsequently, high field magnetisation measurements were carried out (by JGP) at
the Grenoble High Magnetic Field Laboratory, and magnetostriction measurements at the National High Magnetic Field Laboratory, Tallahassee.
The magnetostriction was measured using a miniature capacitance dilatometer~\cite{RM+98} in which the single crystal samples were mounted
with either the $x$, $y$ or $z$ faces parallel to the capacitor plates. The dilatometer could be rotated so that the magnetic field is
perpendicular to the capacitor plate allowing the transverse components of magnetostriction to be measured. 

\begin{figure}
  \begin{center}
    \includegraphics[width=1.00\columnwidth,viewport=0 97 603 705]{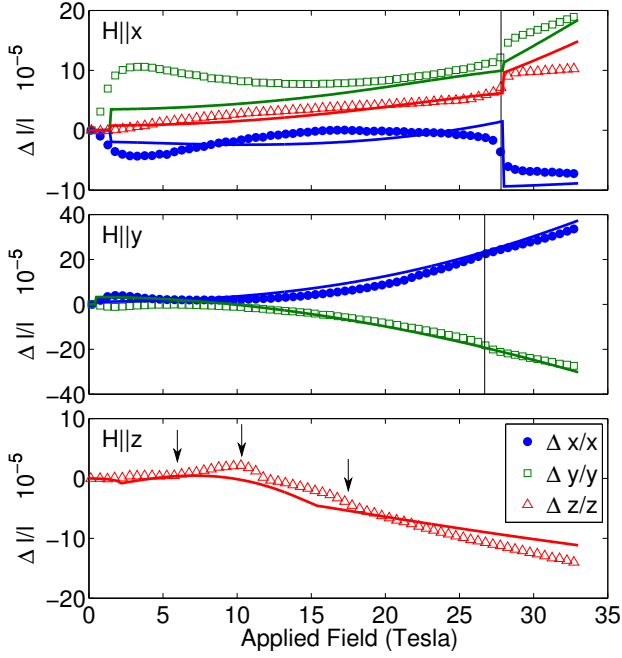}
    \caption{(Color Online) \emph{The high field longitudinal and transverse magnetostriction of UPd$_3$ at 4.2~K.} Solid vertical lines
	     indicate the observed high field transitions. Blue circles, green squares and red triangles denote the magnetostriction
	     parallel to the $x$-, $y$- and $z$-directions, respectively, for the indicated field directions. Solid lines are the results of
             the mean-field calculations.} \label{fg:nhmfl-ms} 
  \end{center}
\end{figure}

Figure~\ref{fg:highfield-mag} shows the magnetisation at several different applied fields, whilst figure~\ref{fg:phasediagram} collates this
and other data~\cite{ZM94} to construct the magnetic phase diagrams of UPd$_3$. The data shows the $T_0$ transition increasing in
temperature with increasing field which is characteristic of an antiferroquadrupolar transition. In general, three ordered phases can be
identified from the data, as the phase between $T_{-1}$ and $T_{+1}$ cannot be distinguished from the magnetisation data.

Figure~\ref{fg:nhmfl-ms} shows the forced magnetostriction data at 4.2 K, plotted as $\Delta l/l$=$[l(H,T=4.2K)-l(H=0,T=4.2K)]/l(H=0,T=4.2K)$.
We measured the longitudinal components of magnetostriction with the field in the $x$-, $y$- and $z$-directions, and also the transverse
components $\Delta y/y$ and $\Delta z/z$ with applied field parallel to $x$, and $\Delta x/x$ with field parallel to $y$. The measurements
were repeated to confirm the reproducibility of the data, and subsequently binned and averaged. In addition, the signal was corrected for
artefacts due to eddy currents. The mechanical noise from the magnet cooling system and electrical noise in the leads meant that we obtained
a resolution of 10$^{-6}$ in $\Delta l/l$.

For $H||x$, the magnetostriction parallel (perpendicular) to the applied field first decreases (increases) until approximately 3 T, then
increases (decreases) to about 15 T before decreasing (increasing) slightly. Similar, but less pronounced, behaviour is also observed
for $H||y$.

Transitions at high field were observed when the field was applied in the basal plane, with a slight anisotropy between the $x$- and
$y$-directions. With applied field parallel to the $x$-direction we see a step-like change in both the longitudinal and transverse
magnetostriction at 28 T, whereas for field parallel to the $y$-direction a change of slope is observed at 27 T. For both field directions,
the magnetostriction parallel to the field decreases (the crystal contracts) at the phase transition, whilst the magnetostriction
perpendicular to the field increases (the crystal expands). 

When the field is applied parallel to the $z$-direction, we observed no high field transitions above 20 T, but instead see anomalies
(indicated by arrows in figure~\ref{fg:nhmfl-ms}) in the longitudinal magnetostriction, at 7~T, 11~T, and 17~T, in agreement with our
magnetisation data and the phase diagram of Tokiwa et al.~\cite{tokiwa01}.

\section{Quadrupolar two-ion interactions and mean field model for UP\lowercase{d}$_3$} \label{sec-upd3-mfcub}

The ordered quadrupolar structures were calculated from a mean field model, with quadrupolar interactions between the quasi-cubic site ions,
using the package \emph{McPhase}~\cite{rotter04,mcphasewebsite}. In order to determine the stable ordered structure, a set of supercells and
corresponding wave vectors is generated. From the wave vector a configuration of moments (dipole, quadrupole etc) is generated and used as
an initial configuration for a self consistent mean field calculation. For each solution of the mean field iteration the free energy is
calculated. The self consistent ordered structure with the lowest free energy is taken to be stable and used for the computation of the
physical properties. In this way the phase boundaries between the different quadrupolar ordered structures were determined in order to
construct the magnetic phase diagrams of UPd$_3$.

The Hamiltonian:

\begin{widetext}
\begin{multline} \label{eq:mfhmltn}
\mathcal{H} = \sum_i \left\{ \mathcal{H}_{\mathrm{cf}}^i + \mathcal{H}_Z^i \right\}
 - \frac{1}{2} \left\{ 
      \mathcal{J}_{11}^{\ij} \sum_{ij}^{\ij} \left[ \hat{J}_{x}^{i} \hat{J}_{x}^{j} + \hat{J}_{y}^{i}\hat{J}_{y}^{j} \right] 
    + \mathcal{K}_{11}^{\ij} \sum_{ij}^{\ij} \left[ \cos(2\phi_{ij})\left(\hat{J}_{x}^{i}\hat{J}_{x}^{j} - \hat{J}_{y}^{i}\hat{J}_{y}^{j} \right)
                                                  + \sin(2\phi_{ij})\left(\hat{J}_{x}^{i}\hat{J}_{y}^{j} + \hat{J}_{y}^{i}\hat{J}_{x}^{j}\right) \right]
    + \mathcal{J}_{10}^{\ij} \sum_{ij}^{\ij} \hat{J}_{z}^{i}\hat{J}_{z}^{j}  \right. \\
    + \mathcal{J}_{21}^{\ij} \sum_{ij}^{\ij} \left[ \hat{Q}_{zx}^{i} \hat{Q}_{zx}^{j} + \hat{Q}_{yz}^{i}\hat{Q}_{yz}^{j} \right]
    + \mathcal{K}_{21}^{\ij} \sum_{ij}^{\ij} \left[ \cos(2\phi_{ij})\left(\hat{Q}_{zx}^{i}\hat{Q}_{zx}^{j} - \hat{Q}_{yz}^{i}\hat{Q}_{yz}^{j} \right)
                                                  + \sin(2\phi_{ij})\left(\hat{Q}_{zx}^{i}\hat{Q}_{yz}^{j} + \hat{Q}_{yz}^{i}\hat{Q}_{xz}^{j}\right) \right] \\ \left.
    + \mathcal{J}_{22}^{\ij} \sum_{ij}^{\ij} \left[ \hat{Q}_{xy}^{i} \hat{Q}_{xy}^{j} + \hat{Q}_{x^2-y^2}^{i}\hat{Q}_{x^2-y^2}^{j} \right] 
    + \mathcal{K}_{22}^{\ij} \sum_{ij}^{\ij} \left[ \cos(4\phi_{ij})\left(\hat{Q}_{xy}^{i}\hat{Q}_{xy}^{j} - \hat{Q}_{x^2-y^2}^{i}\hat{Q}_{x^2-y^2}^{j} \right)
                                                  + \sin(4\phi_{ij})\left(\hat{Q}_{xy}^{i}\hat{Q}_{x^2-y^2}^{j} + \hat{Q}_{x^2-y^2}^{i}\hat{Q}_{xy}^{j}\right) \right]
\right\}
\end{multline}
\end{widetext}

\noindent was employed, where the site indices $i$ and $j$ run over nearest- ($\ij=\nn$) and next-nearest neighbours ($\ij=\nnn$) within an
$ab$ plane and nearest- ($\ij=\nnc$) and next-nearest neighbours ($\ij=\ntc$) between planes. $\mathcal{H}_{\mathrm{cf}}^i$ is the crystal
field Hamiltonian of the $i^{\mathrm{th}}$ quasi-cubic ion given in equation~\ref{eq:Hcfcub} and $\mathcal{H}_Z^i$ is the Zeeman
Hamiltonian, $-g_J\mu_B\V{J}_i\cdot\V{H}$. The form of the two-ion exchange Hamiltonian was derived by considering isotropic interactions
between each pair of neighbours in a local coordinate system defined by an $x'$ axis along the bond, and $z'$ axis along $c$, and then
rotating them into a global coordinate system~\cite{cowleyjensener}. In this way, the full hexagonal symmetry of the interactions is
satisfied. Such expressions were used to explain the properties of elemental Pr~\cite{jensen}, and the dispersion of crystal field
excitations due to interactions between the hexagonal sites in UPd$_3$~\cite{upd3ins}.

The Hamiltonian~(\ref{eq:mfhmltn}) suffices to describe the ordered phases of UPd$_3$. In particular, the anisotropic exchange terms
($\propto\mathcal{K}$) are required since the first (second) order isotropic quadrupolar interactions couple the $\hat{Q}_{zx}$
($\hat{Q}_{xy}$) and $\hat{Q}_{yz}$ ($\hat{Q}_{x^2-y^2}$) operators equally, but the principal order parameters measured by resonant X-ray
diffraction are $\hat{Q}_{zx}$ and $\hat{Q}_{xy}$. Thus, the anisotropic terms are required to favour these ($\hat{Q}_{zx}$, $\hat{Q}_{xy}$)
interactions over the others ($\hat{Q}_{yz}$, $\hat{Q}_{x^2-y^2}$). 

\begin{table*} 
{\large
\begin{center}
  \begin{tabular}{@{\extracolsep{\fill}}rl|rl|rl|rl} \hline
   $\clJ_{10}^{\nn}$  & -0.017    &                    &           & $\clJ_{10}^{\nnc}$  & -0.04     &                     &            \\
   $\clJ_{11}^{\nn}$  & -0.01     & $\clK_{11}^{\nn}$  & -0.02     & $\clJ_{11}^{\nnc}$  & -0.01     & $\clK_{11}^{\nnc}$  &\ 0         \\ \hline
   $\clJ_{21}^{\nn}$  &\ 0.01827  & $\clK_{21}^{\nn}$  &\ 0.00107  & $\clJ_{21}^{\nnc}$  & -0.06088  & $\clK_{21}^{\nnc}$  & -0.00583   \\
   $\clJ_{21}^{\nnn}$ & -1.16\tmv & $\clK_{21}^{\nnn}$ &\ 0.00778  & $\clJ_{21}^{\ntc}$  & -0.00973  & $\clK_{21}^{\ntc}$  & -2.25\tmf  \\ \hline
   $\clJ_{22}^{\nn}$  & -8.42\tmf & $\clK_{22}^{\nn}$  & -3.65\tmf & $\clJ_{22}^{\nnc}$  & -2.55\tmv & $\clK_{22}^{\nnc}$  &\ 0.52\tmv  \\
   $\clJ_{22}^{\nnn}$ & -0.00348  & $\clK_{22}^{\nnn}$ &\ 0.00173  & $\clJ_{22}^{\ntc}$  &\ 0.00318  & $\clK_{22}^{\ntc}$  &\ 2.12\tmf  \\ \hline
  \end{tabular}
  \caption{\emph{Deduced exchange parameters in meV}.}
  \label{tab:upd3expars}
\end{center}
}
\end{table*}

\begin{figure}
  \begin{center}
    \includegraphics[width=0.95\columnwidth,viewport=7 19 584 465]{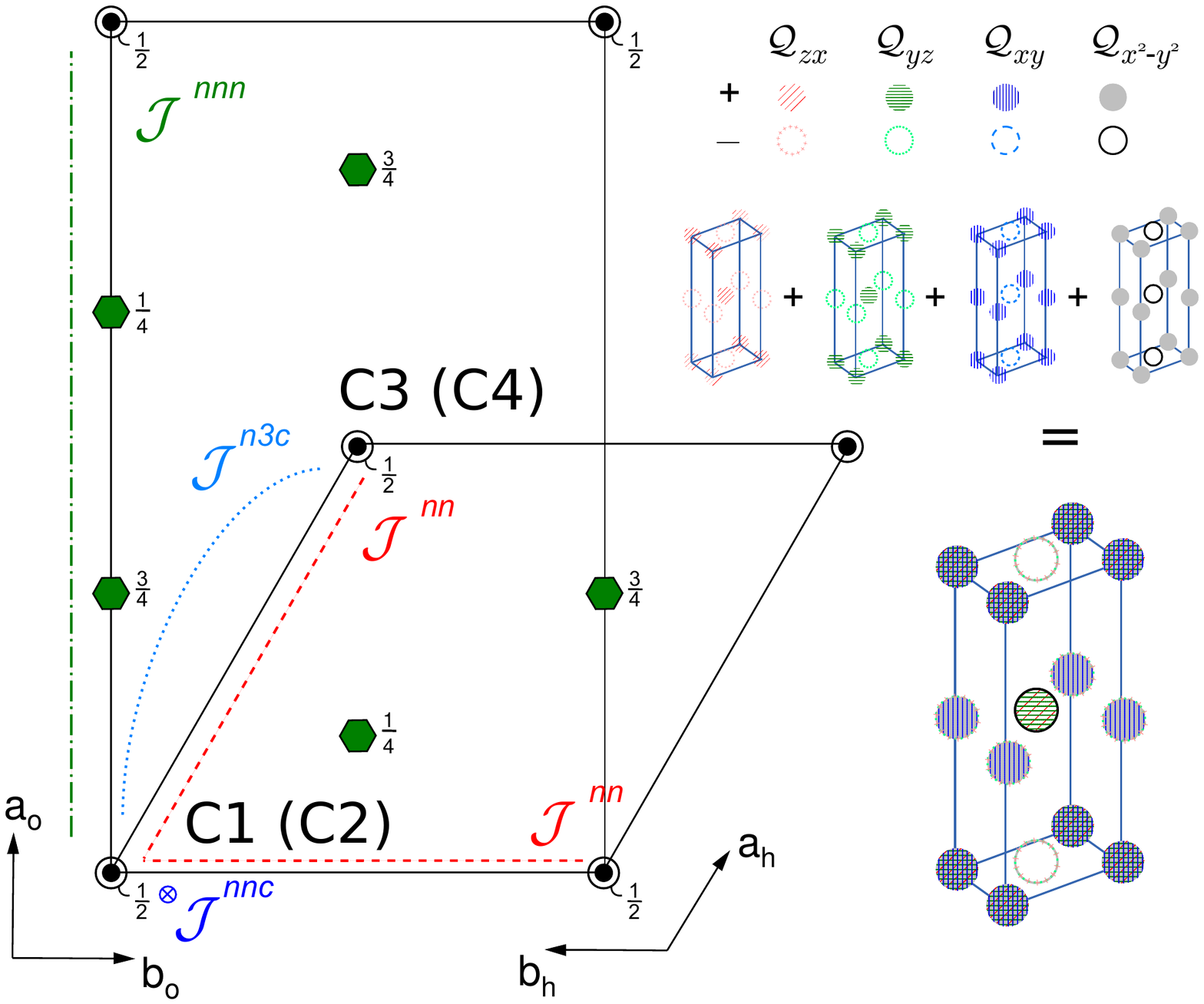}
    \caption{(Color Online) \emph{The double-hexagonal close packed structure.} Only U atoms are shown, each of which has 6-fold coordination 
             with neighbouring Pd-atoms. Circles indicate the quasi-cubic, whilst hexagons the hexagonal sites. Labels in brackets
	     apply to atoms at $z=\sfrac{1}{2}$. Dashed lines show the exchange interaction pathways. Figures on the right illustrate the
	     combination of order parameters which lead to a different quadrupolar moments on each quasi-cubic site in the unit cell, and
             thus to four transitions observed by inelastic neutron scattering at 2~K, as explained in the text.}
    \label{fg:upd3_structure} \end{center}
\end{figure}


In order to stabilise a structure with anti-phase (in-phase) stacking along $c$ we require $\mathcal{J}^{\nnc} < 0$ ($\mathcal{J}^{nnc} > 0$), whilst the AFQ
order in the $ab$ plane requires $\mathcal{J}^{\nn,\nnn} < 0$. Thus, 

\begin{gather} \label{eq:cond1}
\mbox{\small $\mathcal{J}_{zx}^{\nnc} = \mathcal{J}_{21}^{\nnc}+\mathcal{K}_{21}^{\nnc} < 0$ }, \ 
\mbox{\small $\mathcal{J}_{xy}^{\nnc} = \mathcal{J}_{22}^{\nnc}+\mathcal{K}_{22}^{\nnc} > 0$ }, \ 
\mbox{\small $|\mathcal{J}_{zx}^{\nnc}| > |\mathcal{J}_{xy}^{\nnc}|$ } \\ \label{eq:cond2}
 \left[\mathcal{J}_{22}^{\nn}+\mathcal{K}_{22}^{\nn}\cos4\phi_{\nn}\right] < \left[\mathcal{J}_{21}^{\nn}+\mathcal{K}_{21}^{\nn}\cos2\phi_{\nn}\right] < 0
\end{gather}

\noindent should be satisfied in order to yield two phases with $\hat{Q}_{zx}$ anti-phase ($\hat{Q}_{xy}$ in-phase) order along $c$ at
higher (lower) temperatures, as measured. Furthermore, because the $\hat{Q}_{zx}$ ($\hat{Q}_{xy}$) and $\hat{Q}_{x^2-y^2}$ ($\hat{Q}_{yz}$)
operators share the same symmetry~\cite{MPGG03}, an ordering of one of these pairs will induce a secondary ordering of the other quadrupole
of the pair on the same site. That is, a non-zero expectation value $\expbk{\hat{Q}_{zx}}$ implies $\expbk{\hat{Q}_{x^2-y^2}}\neq 0$ also
(angled brackets denote the thermal expectation values $\expbk{\hat{O}}=\sum_n \bra{n}\hat{O}\ket{n}\exp(\frac{-E_n}{k_BT}) / Z$ where the
states $\ket{n}$ are eigenstates of the Hamiltonian~(\ref{eq:mfhmltn}) and $Z$ is the partition function).  This contributes to the
effective field acting on an ion and combines with the exchange interaction to reinforce (or suppress) the ordering of some particular
quadrupole. The strength of this contribution is dependent on the crystal field wavefunctions, and we found that for UPd$_3$, the
$\expbk{\hat{Q}_{x^2-y^2}}$ moment induced by a $\hat{Q}_{zx}$ order is of the same order of magnitude as the primary $\expbk{\hat{Q}_{zx}}$
moment and acts to reinforce the $\hat{Q}_{x^2-y^2}$ ordering (induced $\expbk{\hat{Q}_{x^2-y^2}}$>0). This means that, unfortunately, in
our model there is always a large $\hat{Q}_{x^2-y^2}$ moment in disagreement with the measured XRS azimuthal dependence, which indicates a
10-15\% contribution. 

The other induced moments are generally an order of magnitude weaker than their primary order parameter. They are needed, though, to account
for the observation by inelastic neutron scattering of four almost dispersionless excitations below 4~meV at 1.8~K, which arise from
transitions between the levels of the ground state doublet on the quasi-cubic sites, whose degeneracy is lifted by the quadrupolar order. As
there are four quasi-cubic sites in the ordered unit cell, this implies that the splitting on each site is different, which may only occur
if the magnitude of the quadrupolar moments on each site is different. In-phase ordering of the quadrupoles along $c$ (denoted CS for
$c$-same in the following for brevity) means that the moments on sites C1-C4 of figure~\ref{fg:upd3_structure} are
$\expbk{\hat{Q}^{\mathrm{C1}}}=\expbk{\hat{Q}^{\mathrm{C2}}}=-\expbk{\hat{Q}^{\mathrm{C3}}}=-\expbk{\hat{Q}^{\mathrm{C4}}}$,
whilst anti-phase ordering (henceforth denoted CD, $c$-different) implies that
$\expbk{\hat{Q}^{\mathrm{C1}}}=-\expbk{\hat{Q}^{\mathrm{C2}}}=-\expbk{\hat{Q}^{\mathrm{C3}}}=\expbk{\hat{Q}^{\mathrm{C4}}}$. However, since
$|\expbk{\hat{Q}_{xy}}|=|\expbk{\hat{Q}_{x^2-y^2}}|$ and $|\expbk{\hat{Q}_{zx}}|=|\expbk{\hat{Q}_{yz}}|$~\cite{MPGG03}, the combination of
$\hat{Q}_{zx}$ CD, $\hat{Q}_{yz}$ CD, $\hat{Q}_{xy}$ CS, $\hat{Q}_{x^2-y^2}$ CS ordering imposed by the Hamiltonian~\ref{eq:mfhmltn} 
and conditions~\ref{eq:cond1} and~\ref{eq:cond2} will
result in $\expbk{\hat{Q}^{\mathrm{C1}}}=-\expbk{\hat{Q}^{\mathrm{C3}}}=\expbk{\hat{Q}_{xy}}$ and 
$\expbk{\hat{Q}^{\mathrm{C2}}}=-\expbk{\hat{Q}^{\mathrm{C4}}}=\expbk{\hat{Q}_{zx}}$ so that only two excitations may be expected. Only by
including the induced moments, which yields $\hat{Q}_{x^2-y^2}$ CD and $\hat{Q}_{yz}$ CS ordering (amongst others), will the moments on each
of the sites C1-C4, and thus the splitting of the ground state doublet, be unique. This is illustrated schematically on the right side of
figure~\ref{fg:upd3_structure}.

Finally, we calculate that for the CF scheme outlined in section~\ref{sec-upd3-cf}, $\bra{d_{1,2}}\hat{Q}_{zx}\ket{s}$ and
$\bra{d_{1,2}}\hat{Q}_{yz}\ket{s}$ are an order of magnitude lower than $\bra{d_{1,2}}\hat{Q}_{xy}\ket{s}$ and
$\bra{d_{1,2}}\hat{Q}_{x^2-y^2}\ket{s}$ which means that the exchange parameters $\mathcal{J}_{21}$ should be an order of magnitude larger
than $\mathcal{J}_{22}$ to give similar ordering temperatures for the first and second order quadrupoles as observed~\cite{note1}. From the
above considerations, we first determined the order of magnitude of exchange parameters which result in ordering temperatures below 10~K.
Subsequently a simulated annealing and particle swarm optimisation~\cite{pso} search was carried out to find sets of parameters which yield
at least two transitions at $\approx$4~K and $\approx$8~K and a splitting of the ground state doublet at 2~K close to the measured values
1.28~meV, 1.68~meV, 2.20~meV and 2.60~meV. Additional criteria, including the requirement that the calculated magnetisation with an in-plane
applied field should increase with decreasing temperature, and that there should be non-zero moments of all quadrupoles in the lowest
temperature phases were then used to sort the candidate sets of parameters found by the search. We then calculated the magnetic phase
diagram for the four best sets of parameters, and selected the set which have phase boundaries most similar to those measured. Finally, the
parameters were refined by hand to better match the transition temperatures and fields.

We found during this procedure that unless the dipolar interactions are included, structures with ordering wavevectors
$(\frac{1}{2}\frac{1}{2}0)$ are favoured over the $(\frac{1}{2}00)$ observed (indexed with respect to the dhcp cell). Thus small values of
$\mathcal{J}_{1m}$ and $\mathcal{K}_{1m}$ were included, but not varied in the search procedure. They were subsequently refined by hand
along with the quadrupolar interaction parameters to better fit the measured critical fields.

The final parameters are shown in table~\ref{tab:upd3expars}. At 2~K, the \emph{full} mean field model with these parameters yields a
splitting of the doublet ground state on ions C1-C4 of 0.69~meV, 1.65~meV, 1.82~meV, 2.08~meV respectively (at
$Q=(\frac{2}{3}\frac{2}{3}0)$), which is somewhat lower than the experimentally measured values (1.28~meV, 1.68~meV, 2.20~meV and
2.60~meV~\cite{MSM93}). The dispersion of the levels along $[00l]$, $\approx0.5$~meV, is close to the measured value but the in plane
dispersion of $\approx1$~meV is in stark contrast to the measurements which showed the modes to be almost dispersionless.

Furthermore, the calculated order parameters differ from the XRS measurements: in the model, the dominant order parameters are
$\hat{Q}_{x^2-y^2}$ and $\hat{Q}_{xy}$ rather than $\hat{Q}_{zx}$ and $\hat{Q}_{xy}$. The phase denoted AFQ1 in figure~\ref{fg:phasediagram}
has a large $\hat{Q}_{x^2-y^2}$ moment (ordered in antiphase along $c$) inducing a small ($\approx$ 2\%) $\hat{Q}_{zx}$ moment. In the AFQ2
phase, $\hat{Q}_{xy}$ moments become ordered, inducing some $\hat{Q}_{yz}$ quadrupoles; both these quadrupolar moments double below the
transition to the AFQ3 phase. Thus although the calculated sequence of ordering agrees with experimental data, the type of quadrupolar order
does not. Unfortunately, we found it impossible to stabilise the $\hat{Q}_{zx}$ order parameter over the $\hat{Q}_{x^2-y^2}$ order parameter
at higher temperatures whilst maintaining a ground state with all quadrupolar moments ordered.

Another discrepancy between the model and experiments is the very low critical fields when $H||c$ (seen at the bottom of figure~\ref{fg:phasediagram})
and the magnitude of the $c$ axis magnetisation which is some three times smaller than measured (figure~\ref{fg:highfield-mag}). In principle,
this can be altered by increasing the $\mathcal{J}_{10}$ exchange parameters, however, we found that raising these from the values in
table~\ref{tab:upd3expars} suppresses the quadrupolar ordering completely, in favour of a dipolar order. Alternatively, the quadrupolar exchange 
parameters $\mathcal{J}_{2m}$ may be altered, but an increase in the critical field necessitates also an increase in the transition temperatures.

Since the exchange interactions are likely to arise from the RKKY mechanism, one expects that it should be long ranged. Thus including
interactions further than nearest neighbour may give better agreement with the data. However, this vastly increases the parameter space, and
unfortunately we could not obtain good fits with simple analytical forms of the RKKY exchange. A more sophisticated approach, using the
measured bandstructure of the Pd-U conduction band, may give better results. This consideration may also apply to the quadrupolar
interaction parameters, and may account for the discrepancies between the measured and calculated phase boundaries.

Finally, we note that ultrasound measurements~\cite{LMMM99} and a symmetry analysis of the XRS data~\cite{lovesey2010} showed that there is
a sequence of structural transitions from hexagonal to orthorhombic symmetry at $T_0$ and from orthorhombic to monoclinic at $T_{+1}$.  As
the parameters in table~\ref{tab:upd3expars} have the symmetry of the high temperature hexagonal structure, it is possible that using
temperature dependent exchange parameters which incorporate deviations from hexagonal symmetry proportional to the order parameter may yield
a better fit.


The magnetoelastic strain is proportional to the strain derivative of the free energy~\cite{rotterdoerrlindbaum}, 

\begin{equation} \label{eq:epsilon}
\epsilon^\alpha = -\sum_\beta \frac{s^{\alpha\beta}}{V} \frac{\partial F}{\partial \epsilon^\beta}
\end{equation}

\noindent where $s^{\alpha\beta}$ is the elastic compliance and the indices $\alpha$ and $\beta$ are the Cartesian directions. Noting that $F=-k_B
T\ln Z$, where the partition function is $Z=\sum_n \exp(-E_n/k_B T)$, we thus find that $\frac{\partial F}{\partial \epsilon^\beta}=\langle
\frac{\partial E_n}{\partial \epsilon^\beta}\rangle$ where the angled brackets indicate the thermal expectation value.

In the context of the Hamiltonian developed above, there are two main contributions to the magnetoelastic strain. These arise from the single-ion crystal 
field and the two-ion exchange interactions, both of which depend on the position of the ions, and are hence coupled to any change in the lattice. These two
interactions give rise to \emph{crystal field striction}~\cite{MS90} and \emph{exchange striction}~\cite{rotterdoerrlindbaum}, respectively. In order
to calculate the magnetostriction, we thus have to find expressions for the energy levels $E_n$ as a function of these two interactions. This is done
by expanding the Hamiltonian, equation~\ref{eq:mfhmltn}, in a Taylor series and making the \emph{harmonic approximation} by keeping only the first
order term, which yields

\begin{eqnarray} \label{eq:cfstrain}
\epsilon^\alpha_{\mathrm{cf}} &=& - \sum_{kq,i} \frac{s^{\alpha\beta}}{V} \frac{\partial B_k^q}{\partial\epsilon} \langle O_k^q \rangle \\ \label{eq:exstrain}
\epsilon^\alpha_{\mathrm{ex}} &=& \frac{1}{2} \sum_{\beta,ij}   \frac{s^{\alpha\beta}}{V} \frac{\partial \mathcal{J}_{ij}}{\partial\epsilon^\beta} 
                                    \langle \hat{J}_i^{\beta} \hat{J}_j^{\beta} \rangle
\end{eqnarray}

\noindent The prefactors $A_{\alpha}=\frac{s^{\alpha\beta}}{V} \frac{\partial B_q^k}{\partial\epsilon}$ and $K_{\alpha\beta}^{ij}=\frac{s^{\alpha\beta}}{V}
\frac{\partial \mathcal{J}_{ij}}{\partial\epsilon^\beta}$ may be taken to be independent of field and temperature and can then be fitted to experimental 
data given the thermal expectation values $\langle O_k^q \rangle$ and $\langle \hat{J}_i \hat{J}_j \rangle$ obtained from the mean-field model at different
applied magnetic fields.

Considering only the non-zero terms up to rank-2 in the Hamiltonian (\ref{eq:mfhmltn}), we thus have:

\begin{multline}
\label{eq:ms}
\epsilon^{\alpha} = \frac{1}{N} \sum_i \left[ \sum_{\beta} A_{\alpha\beta} \langle \hat{Q}_{\beta}^i \rangle
        + \sum_{\beta} K_{\alpha\beta}^{ab} \langle \hat{Q}_{\beta}^i \hat{Q}_{\beta}^{i+(\V{a}+\V{b})/2} \rangle
\right. \\ \left.
        + \sum_{\beta} K_{\alpha\beta}^{c} \langle \hat{Q}_{\beta}^i \hat{Q}_{\beta}^{i+\V{c}} \rangle \right]
\end{multline}

\noindent where $N=4$ is the number of U$^{4+}$ ions in the magnetic unit cell, and the indices $\alpha=x,y,z$,
$\beta=xy,yz,z^2,zx,x^2-y^2$. In order to reduce the number of parameters in fitting equation~\ref{eq:ms} to the data, we considered only
the quadrupolar exchange interactions between nearest neighbour ions in the $c$-direction and the basal plane, since the calculated
expectation values of $\langle \hat{Q}_i \hat{Q}_j \rangle \gg \langle \hat{J}_i \hat{J}_j \rangle$. Table~\ref{tab:upd3ms} shows the
fitted parameters, whilst the calculated magnetostriction is shown as solid lines in figure~\ref{fg:nhmfl-ms}.

\begin{table} 
\begin{center}
  \begin{tabular}{r|c|c|c} \hline
                                                    &           \multicolumn{3}{c}{$\alpha$}              \\ \cline{2-4}
                                                    &       $a$       &       $b$       &       $c$       \\ \hline
    $A_{\alpha,xy}\times 10^{-5}$                   & ~~~~~ 26  ~~~~  &~~~~~~-100 ~~~~  & ~~~~      ~~~~  \\
    $A_{\alpha,yz}\times 10^{-5}$                   & ~~~~ -35  ~~~~  & ~~~~~ 85  ~~~~  & ~~~~      ~~~~  \\
    $A_{\alpha,z^2}\times 10^{-5}$                  & ~~~~ -1 ~~~~~~  & ~~~~~ 5 ~~~~~~  & ~~~~ -30 ~~~~~  \\
    $A_{\alpha,zx}\times 10^{-5}$                   & ~~~~~ 80  ~~~~  & ~~~~ -50  ~~~~  & ~~~~ -30 ~~~~~  \\
    $A_{\alpha,x^2-y^2}\times 10^{-5}$              & ~~~~  5  ~~~~~  & ~~~~ -60  ~~~~  & ~~~~ -30 ~~~~~  \\ \hline
    $K_{\alpha,xy}^{ab}\times 10^{-5}$              & ~~~~~ 48  ~~~~  & ~~~~~ 20  ~~~~  & ~~~~      ~~~~  \\
    $K_{\alpha,yz}^{ab}\times 10^{-5}$              &~~~~~ -10 ~~~~~  & ~~~~~ 30  ~~~~  & ~~~~      ~~~~  \\
    $K_{\alpha,zx}^{ab}\times 10^{-5}$              &~~~~~~-150  ~~~  & ~~~~ -45  ~~~~  & ~~~~  25  ~~~~  \\
    $K_{\alpha,x^2-y^2}^{ab}\times 10^{-5}$         & ~~~~  1  ~~~~~  & ~~~~~ 10  ~~~~  & ~~~~  10  ~~~~  \\ \hline
    $K_{\alpha,xy}^{c}\times 10^{-5}$               & ~~~~ -48  ~~~~  & ~~~~ -14  ~~~~  & ~~~~      ~~~~  \\
    $K_{\alpha,yz}^{c}\times 10^{-5}$               & ~~~~ -3 ~~~~~~  & ~~~~ -40  ~~~~  & ~~~~  5 ~~~~~~  \\
    $K_{\alpha,zx}^{c}\times 10^{-5}$               &~~~~~~ 200  ~~~  & ~~~~~ 35  ~~~~  & ~~~~      ~~~~  \\
    $K_{\alpha,x^2-y^2}^{c}\times 10^{-5}$          & ~~~~  8  ~~~~~  & ~~~~ -4 ~~~~~~  & ~~~~      ~~~~  \\ \hline
  \end{tabular}
  \caption{\emph{Fitted magnetoelastic parameters}.}
  \label{tab:upd3ms}
\end{center}
\end{table}

The calculated magnetostriction fits the data well in the high field regions, but does not reproduce the low field behaviour, particular for the case
where the field is applied along $x$. It may be that at low fields, some of the measured magnetostriction is due to domain rotation, which is not
considered in the mean field model.

\section{Conclusions} \label{sec-upd3-conc}

We have deduced a mean field model including up to four nearest neighbour dipolar exchange and quadrupolar interactions between the 5$f^2$
electrons of UPd$_3$ which is in generally good qualitative agreement with a broad range of experimental results. The interactions between
electrons on the quasi-cubic sites were deduced from resonant x-ray scattering measurements of the quadrupolar order parameters of each of
the four low temperature ordered phases, and from the measured transition temperatures and critical fields. The model was then used to
calculate the high field magnetisation and magnetostriction, and the magnetic phase diagram up to 30 T. With an applied magnetic field in
the basal plane, the calculations for the $x$ and $y$ directions agree with the data. The calculated magnetisation with the field parallel
to $z$ is a factor of 3 too large, however, and the calculated critical field in this direction is much smaller than that measured. Despite
this, the $z$-direction calculated magnetic phase diagrams qualitatively reproduces that measured. In conclusion, we have developed a
nearest neighbour mean-field model which reproduces many of the main features observed in UPd$_3$. 

\section*{Acknowledgements}

The authors thank Jens Jensen, Gillian Gehring, Amanda Gipson and Helen Walker for fruitful discussions. M.D.L. thanks the UK Engineering
and Physical Sciences Research Council for a research studentship, and UCL Graduate School for a project grant. J.G.P. acknowledges
assistance by G. Chouteau and the support provided by the Grenoble High Magnetic Field Laboratory. M.D.L. and M.R. acknowledge the support
of the European Science Foundation's Cooperation in Science and Technology (COST) program, contract number COST-STSM-P16-02275. Experiments
performed at the NHMFL were supported by NSF Cooperative Agreement DMR-0084173, by the state of Florida, and the U.S. Department of Energy.
Support by the Deustche Forschungsgemeinschaft (SFB463), and the Austrian FWF (P17226) is gratefully acknowledged. 




\begin{thebibliography}{35}%
\makeatletter
\providecommand \@ifxundefined [1]{%
 \@ifx{#1\undefined}
}%
\providecommand \@ifnum [1]{%
 \ifnum #1\expandafter \@firstoftwo
 \else \expandafter \@secondoftwo
 \fi
}%
\providecommand \@ifx [1]{%
 \ifx #1\expandafter \@firstoftwo
 \else \expandafter \@secondoftwo
 \fi
}%
\providecommand \natexlab [1]{#1}%
\providecommand \enquote  [1]{``#1''}%
\providecommand \bibnamefont  [1]{#1}%
\providecommand \bibfnamefont [1]{#1}%
\providecommand \citenamefont [1]{#1}%
\providecommand \href@noop [0]{\@secondoftwo}%
\providecommand \href [0]{\begingroup \@sanitize@url \@href}%
\providecommand \@href[1]{\@@startlink{#1}\@@href}%
\providecommand \@@href[1]{\endgroup#1\@@endlink}%
\providecommand \@sanitize@url [0]{\catcode `\\12\catcode `\$12\catcode
  `\&12\catcode `\#12\catcode `\^12\catcode `\_12\catcode `\%12\relax}%
\providecommand \@@startlink[1]{}%
\providecommand \@@endlink[0]{}%
\providecommand \url  [0]{\begingroup\@sanitize@url \@url }%
\providecommand \@url [1]{\endgroup\@href {#1}{\urlprefix }}%
\providecommand \urlprefix  [0]{URL }%
\providecommand \Eprint [0]{\href }%
\providecommand \doibase [0]{http://dx.doi.org/}%
\providecommand \selectlanguage [0]{\@gobble}%
\providecommand \bibinfo  [0]{\@secondoftwo}%
\providecommand \bibfield  [0]{\@secondoftwo}%
\providecommand \translation [1]{[#1]}%
\providecommand \BibitemOpen [0]{}%
\providecommand \bibitemStop [0]{}%
\providecommand \bibitemNoStop [0]{.\EOS\space}%
\providecommand \EOS [0]{\spacefactor3000\relax}%
\providecommand \BibitemShut  [1]{\csname bibitem#1\endcsname}%
\let\auto@bib@innerbib\@empty
\bibitem [{\citenamefont {Santini}\ \emph {et~al.}(2009)\citenamefont
  {Santini}, \citenamefont {Carretta}, \citenamefont {Amoretti}, \citenamefont
  {Caciuffo}, \citenamefont {Magnani},\ and\ \citenamefont
  {Lander}}]{rmp_multipoles}%
  \BibitemOpen
  \bibfield  {author} {\bibinfo {author} {\bibfnamefont {P.}~\bibnamefont
  {Santini}}, \bibinfo {author} {\bibfnamefont {S.}~\bibnamefont {Carretta}},
  \bibinfo {author} {\bibfnamefont {G.}~\bibnamefont {Amoretti}}, \bibinfo
  {author} {\bibfnamefont {R.}~\bibnamefont {Caciuffo}}, \bibinfo {author}
  {\bibfnamefont {N.}~\bibnamefont {Magnani}}, \ and\ \bibinfo {author}
  {\bibfnamefont {G.~H.}\ \bibnamefont {Lander}},\ }\href {\doibase
  10.1103/RevModPhys.81.807} {\bibfield  {journal} {\bibinfo  {journal} {Rev.
  Mod. Phys.}\ }\textbf {\bibinfo {volume} {81}},\ \bibinfo {pages} {807}
  (\bibinfo {year} {2009})}\BibitemShut {NoStop}%
\bibitem [{\citenamefont {Yakhou}\ \emph {et~al.}(2001)\citenamefont {Yakhou},
  \citenamefont {Plakhty}, \citenamefont {Suzuki}, \citenamefont {Gavrilov},
  \citenamefont {Burlet}, \citenamefont {Paolasini}, \citenamefont {Vettier},\
  and\ \citenamefont {Kunii}}]{ceb6xrayEU}%
  \BibitemOpen
  \bibfield  {author} {\bibinfo {author} {\bibfnamefont {F.}~\bibnamefont
  {Yakhou}}, \bibinfo {author} {\bibfnamefont {V.}~\bibnamefont {Plakhty}},
  \bibinfo {author} {\bibfnamefont {H.}~\bibnamefont {Suzuki}}, \bibinfo
  {author} {\bibfnamefont {S.}~\bibnamefont {Gavrilov}}, \bibinfo {author}
  {\bibfnamefont {P.}~\bibnamefont {Burlet}}, \bibinfo {author} {\bibfnamefont
  {L.}~\bibnamefont {Paolasini}}, \bibinfo {author} {\bibfnamefont
  {C.}~\bibnamefont {Vettier}}, \ and\ \bibinfo {author} {\bibfnamefont
  {S.}~\bibnamefont {Kunii}},\ }\href {\doibase 10.1016/S0375-9601(01)00338-3}
  {\bibfield  {journal} {\bibinfo  {journal} {Phys. Lett. A}\ }\textbf
  {\bibinfo {volume} {285}},\ \bibinfo {pages} {191} (\bibinfo {year}
  {2001})}\BibitemShut {NoStop}%
\bibitem [{\citenamefont {Nakao}\ \emph {et~al.}(2001)\citenamefont {Nakao},
  \citenamefont {Magishi}, \citenamefont {Wakabayashi}, \citenamefont
  {Murakami}, \citenamefont {Koyama}, \citenamefont {Hirota}, \citenamefont
  {Endoh},\ and\ \citenamefont {Kunii}}]{ceb6xrayjap}%
  \BibitemOpen
  \bibfield  {author} {\bibinfo {author} {\bibfnamefont {H.}~\bibnamefont
  {Nakao}}, \bibinfo {author} {\bibfnamefont {K.}~\bibnamefont {Magishi}},
  \bibinfo {author} {\bibfnamefont {Y.}~\bibnamefont {Wakabayashi}}, \bibinfo
  {author} {\bibfnamefont {Y.}~\bibnamefont {Murakami}}, \bibinfo {author}
  {\bibfnamefont {K.}~\bibnamefont {Koyama}}, \bibinfo {author} {\bibfnamefont
  {K.}~\bibnamefont {Hirota}}, \bibinfo {author} {\bibfnamefont
  {Y.}~\bibnamefont {Endoh}}, \ and\ \bibinfo {author} {\bibfnamefont
  {S.}~\bibnamefont {Kunii}},\ }\href {\doibase 10.1143/JPSJ.70.1857}
  {\bibfield  {journal} {\bibinfo  {journal} {J. Phys. Soc. Japan}\ }\textbf
  {\bibinfo {volume} {70}},\ \bibinfo {pages} {1857} (\bibinfo {year}
  {2001})}\BibitemShut {NoStop}%
\bibitem [{\citenamefont {Onimaru}\ \emph {et~al.}(2005)\citenamefont
  {Onimaru}, \citenamefont {Sakakibara}, \citenamefont {Tayama}, \citenamefont
  {Aso}, \citenamefont {amd Dai~Aoki}, \citenamefont {\={O}nuki}, \citenamefont
  {Kawae}, \citenamefont {Kitai},\ and\ \citenamefont
  {Takeuchi}}]{onimaru_prpb3}%
  \BibitemOpen
  \bibfield  {author} {\bibinfo {author} {\bibfnamefont {T.}~\bibnamefont
  {Onimaru}}, \bibinfo {author} {\bibfnamefont {T.}~\bibnamefont {Sakakibara}},
  \bibinfo {author} {\bibfnamefont {T.}~\bibnamefont {Tayama}}, \bibinfo
  {author} {\bibfnamefont {N.}~\bibnamefont {Aso}}, \bibinfo {author}
  {\bibfnamefont {H.~Y.}\ \bibnamefont {amd Dai~Aoki}}, \bibinfo {author}
  {\bibfnamefont {Y.}~\bibnamefont {\={O}nuki}}, \bibinfo {author}
  {\bibfnamefont {T.}~\bibnamefont {Kawae}}, \bibinfo {author} {\bibfnamefont
  {T.}~\bibnamefont {Kitai}}, \ and\ \bibinfo {author} {\bibfnamefont
  {T.}~\bibnamefont {Takeuchi}},\ }\href {\doibase 10.1016/j.physb.2005.01.268}
  {\bibfield  {journal} {\bibinfo  {journal} {Physica B}\ }\textbf {\bibinfo
  {volume} {359--361}},\ \bibinfo {pages} {935} (\bibinfo {year}
  {2005})}\BibitemShut {NoStop}%
\bibitem [{\citenamefont {Nagao}\ and\ \citenamefont
  {Shiina}(2010)}]{nagao_tmte}%
  \BibitemOpen
  \bibfield  {author} {\bibinfo {author} {\bibfnamefont {T.}~\bibnamefont
  {Nagao}}\ and\ \bibinfo {author} {\bibfnamefont {R.}~\bibnamefont {Shiina}},\
  }\href {\doibase 10.1143/JPSJ.79.034703} {\bibfield  {journal} {\bibinfo
  {journal} {J. Phys. Soc. Japan}\ }\textbf {\bibinfo {volume} {79}},\ \bibinfo
  {pages} {034703} (\bibinfo {year} {2010})}\BibitemShut {NoStop}%
\bibitem [{\citenamefont {Tanaka}\ \emph {et~al.}(2004)\citenamefont {Tanaka},
  \citenamefont {Inami}, \citenamefont {Lovesey}, , \citenamefont {Knight},
  \citenamefont {Yakhou}, \citenamefont {Mannix}, \citenamefont {Kokubun},
  \citenamefont {Kanazawa}, \citenamefont {Ishida}, \citenamefont {Nanao},
  \citenamefont {Nakamura}, \citenamefont {Yamauchi}, \citenamefont {Onodera},
  \citenamefont {Ohoyama},\ and\ \citenamefont {Yamaguchi}}]{tanaka_dyb2c2}%
  \BibitemOpen
  \bibfield  {author} {\bibinfo {author} {\bibfnamefont {Y.}~\bibnamefont
  {Tanaka}}, \bibinfo {author} {\bibfnamefont {T.}~\bibnamefont {Inami}},
  \bibinfo {author} {\bibfnamefont {S.~W.}\ \bibnamefont {Lovesey}}, , \bibinfo
  {author} {\bibfnamefont {K.~S.}\ \bibnamefont {Knight}}, \bibinfo {author}
  {\bibfnamefont {F.}~\bibnamefont {Yakhou}}, \bibinfo {author} {\bibfnamefont
  {D.}~\bibnamefont {Mannix}}, \bibinfo {author} {\bibfnamefont
  {J.}~\bibnamefont {Kokubun}}, \bibinfo {author} {\bibfnamefont
  {M.}~\bibnamefont {Kanazawa}}, \bibinfo {author} {\bibfnamefont
  {K.}~\bibnamefont {Ishida}}, \bibinfo {author} {\bibfnamefont
  {S.}~\bibnamefont {Nanao}}, \bibinfo {author} {\bibfnamefont
  {T.}~\bibnamefont {Nakamura}}, \bibinfo {author} {\bibfnamefont
  {H.}~\bibnamefont {Yamauchi}}, \bibinfo {author} {\bibfnamefont
  {H.}~\bibnamefont {Onodera}}, \bibinfo {author} {\bibfnamefont
  {K.}~\bibnamefont {Ohoyama}}, \ and\ \bibinfo {author} {\bibfnamefont
  {Y.}~\bibnamefont {Yamaguchi}},\ }\href {\doibase 10.1103/PhysRevB.69.024417}
  {\bibfield  {journal} {\bibinfo  {journal} {Phys. Rev. B}\ }\textbf {\bibinfo
  {volume} {69}},\ \bibinfo {pages} {024417} (\bibinfo {year}
  {2004})}\BibitemShut {NoStop}%
\bibitem [{\citenamefont {Walker}\ \emph {et~al.}(2006)\citenamefont {Walker},
  \citenamefont {McEwen}, \citenamefont {McMorrow}, \citenamefont {Wilkins},
  \citenamefont {Wastin}, \citenamefont {Colineau},\ and\ \citenamefont
  {Fort}}]{WMM+06}%
  \BibitemOpen
  \bibfield  {author} {\bibinfo {author} {\bibfnamefont {H.~C.}\ \bibnamefont
  {Walker}}, \bibinfo {author} {\bibfnamefont {K.~A.}\ \bibnamefont {McEwen}},
  \bibinfo {author} {\bibfnamefont {D.~F.}\ \bibnamefont {McMorrow}}, \bibinfo
  {author} {\bibfnamefont {S.~B.}\ \bibnamefont {Wilkins}}, \bibinfo {author}
  {\bibfnamefont {F.}~\bibnamefont {Wastin}}, \bibinfo {author} {\bibfnamefont
  {E.}~\bibnamefont {Colineau}}, \ and\ \bibinfo {author} {\bibfnamefont
  {D.}~\bibnamefont {Fort}},\ }\href {\doibase 10.1103/PhysRevLett.97.137203}
  {\bibfield  {journal} {\bibinfo  {journal} {Phys. Rev. Lett.}\ }\textbf
  {\bibinfo {volume} {97}},\ \bibinfo {pages} {137203} (\bibinfo {year}
  {2006})}\BibitemShut {NoStop}%
\bibitem [{\citenamefont {Walker}\ \emph {et~al.}(2008)\citenamefont {Walker},
  \citenamefont {McEwen}, \citenamefont {Le}, \citenamefont {Paolasini},\ and\
  \citenamefont {Fort}}]{hcwESRF07}%
  \BibitemOpen
  \bibfield  {author} {\bibinfo {author} {\bibfnamefont {H.~C.}\ \bibnamefont
  {Walker}}, \bibinfo {author} {\bibfnamefont {K.~A.}\ \bibnamefont {McEwen}},
  \bibinfo {author} {\bibfnamefont {M.~D.}\ \bibnamefont {Le}}, \bibinfo
  {author} {\bibfnamefont {L.}~\bibnamefont {Paolasini}}, \ and\ \bibinfo
  {author} {\bibfnamefont {D.}~\bibnamefont {Fort}},\ }\href {\doibase
  10.1088/0953-8984/20/39/395221} {\bibfield  {journal} {\bibinfo  {journal}
  {J. Phys.: Condens. Matter}\ }\textbf {\bibinfo {volume} {20}},\ \bibinfo
  {pages} {395221} (\bibinfo {year} {2008})}\BibitemShut {NoStop}%
\bibitem [{\citenamefont {Zochowski}\ \emph {et~al.}(1995)\citenamefont
  {Zochowski}, \citenamefont {de~Podesta}, \citenamefont {Lester},\ and\
  \citenamefont {McEwen}}]{ZdPM95}%
  \BibitemOpen
  \bibfield  {author} {\bibinfo {author} {\bibfnamefont {S.~W.}\ \bibnamefont
  {Zochowski}}, \bibinfo {author} {\bibfnamefont {M.}~\bibnamefont
  {de~Podesta}}, \bibinfo {author} {\bibfnamefont {C.}~\bibnamefont {Lester}},
  \ and\ \bibinfo {author} {\bibfnamefont {K.~A.}\ \bibnamefont {McEwen}},\
  }\href {\doibase 10.1016/0921-4526(94)00499-L} {\bibfield  {journal}
  {\bibinfo  {journal} {Physica B}\ }\textbf {\bibinfo {volume} {206\&207}},\
  \bibinfo {pages} {489} (\bibinfo {year} {1995})}\BibitemShut {NoStop}%
\bibitem [{\citenamefont {McEwen}\ \emph {et~al.}(2003)\citenamefont {McEwen},
  \citenamefont {Park}, \citenamefont {Gipson},\ and\ \citenamefont
  {Gehring}}]{MPGG03}%
  \BibitemOpen
  \bibfield  {author} {\bibinfo {author} {\bibfnamefont {K.~A.}\ \bibnamefont
  {McEwen}}, \bibinfo {author} {\bibfnamefont {J.-G.}\ \bibnamefont {Park}},
  \bibinfo {author} {\bibfnamefont {A.~J.}\ \bibnamefont {Gipson}}, \ and\
  \bibinfo {author} {\bibfnamefont {G.~A.}\ \bibnamefont {Gehring}},\ }\href
  {\doibase 10.1088/0953-8984/15/28/304} {\bibfield  {journal} {\bibinfo
  {journal} {J. Phys: Cond. Mat.}\ }\textbf {\bibinfo {volume} {15}},\ \bibinfo
  {pages} {S1923} (\bibinfo {year} {2003})}\BibitemShut {NoStop}%
\bibitem [{\citenamefont {Tokiwa}\ \emph {et~al.}(2001)\citenamefont {Tokiwa},
  \citenamefont {Sugiyama}, \citenamefont {Takeuchi}, \citenamefont
  {Nakashima}, \citenamefont {Settai}, \citenamefont {Inada}, \citenamefont
  {Haga}, \citenamefont {Yamamoto}, \citenamefont {Kindo}, \citenamefont
  {Harima},\ and\ \citenamefont {Onuki}}]{tokiwa01}%
  \BibitemOpen
  \bibfield  {author} {\bibinfo {author} {\bibfnamefont {Y.}~\bibnamefont
  {Tokiwa}}, \bibinfo {author} {\bibfnamefont {K.}~\bibnamefont {Sugiyama}},
  \bibinfo {author} {\bibfnamefont {T.}~\bibnamefont {Takeuchi}}, \bibinfo
  {author} {\bibfnamefont {M.}~\bibnamefont {Nakashima}}, \bibinfo {author}
  {\bibfnamefont {R.}~\bibnamefont {Settai}}, \bibinfo {author} {\bibfnamefont
  {Y.}~\bibnamefont {Inada}}, \bibinfo {author} {\bibfnamefont
  {Y.}~\bibnamefont {Haga}}, \bibinfo {author} {\bibfnamefont {E.}~\bibnamefont
  {Yamamoto}}, \bibinfo {author} {\bibfnamefont {K.}~\bibnamefont {Kindo}},
  \bibinfo {author} {\bibfnamefont {H.}~\bibnamefont {Harima}}, \ and\ \bibinfo
  {author} {\bibfnamefont {Y.}~\bibnamefont {Onuki}},\ }\href {\doibase
  10.1143/JPSJ.70.1731} {\bibfield  {journal} {\bibinfo  {journal} {J. Phys.
  Soc. Jpn.}\ }\textbf {\bibinfo {volume} {70}},\ \bibinfo {pages} {1731}
  (\bibinfo {year} {2001})}\BibitemShut {NoStop}%
\bibitem [{\citenamefont {Steigenberger}\ \emph {et~al.}(1992)\citenamefont
  {Steigenberger}, \citenamefont {McEwen}, \citenamefont {Martinez},\ and\
  \citenamefont {Fort}}]{SMMF92}%
  \BibitemOpen
  \bibfield  {author} {\bibinfo {author} {\bibfnamefont {U.}~\bibnamefont
  {Steigenberger}}, \bibinfo {author} {\bibfnamefont {K.~A.}\ \bibnamefont
  {McEwen}}, \bibinfo {author} {\bibfnamefont {J.~L.}\ \bibnamefont
  {Martinez}}, \ and\ \bibinfo {author} {\bibfnamefont {D.}~\bibnamefont
  {Fort}},\ }\href {\doibase 10.1016/0304-8853(92)91395-A} {\bibfield
  {journal} {\bibinfo  {journal} {J. Mag. Mag. Mater.}\ }\textbf {\bibinfo
  {volume} {108}},\ \bibinfo {pages} {163} (\bibinfo {year}
  {1992})}\BibitemShut {NoStop}%
\bibitem [{\citenamefont {McMorrow}\ \emph {et~al.}(2001)\citenamefont
  {McMorrow}, \citenamefont {McEwen}, \citenamefont {Steigenberger},
  \citenamefont {R{\o}nnow},\ and\ \citenamefont {Yakhou}}]{MMS+01}%
  \BibitemOpen
  \bibfield  {author} {\bibinfo {author} {\bibfnamefont {D.~F.}\ \bibnamefont
  {McMorrow}}, \bibinfo {author} {\bibfnamefont {K.~A.}\ \bibnamefont
  {McEwen}}, \bibinfo {author} {\bibfnamefont {U.}~\bibnamefont
  {Steigenberger}}, \bibinfo {author} {\bibfnamefont {H.~M.}\ \bibnamefont
  {R{\o}nnow}}, \ and\ \bibinfo {author} {\bibfnamefont {F.}~\bibnamefont
  {Yakhou}},\ }\href {\doibase 10.1103/PhysRevLett.87.057201} {\bibfield
  {journal} {\bibinfo  {journal} {Phys. Rev. Lett.}\ }\textbf {\bibinfo
  {volume} {87}},\ \bibinfo {pages} {057201} (\bibinfo {year}
  {2001})}\BibitemShut {NoStop}%
\bibitem [{\citenamefont {Heal}\ and\ \citenamefont {Williams}(1955)}]{HW55}%
  \BibitemOpen
  \bibfield  {author} {\bibinfo {author} {\bibfnamefont {T.~J.}\ \bibnamefont
  {Heal}}\ and\ \bibinfo {author} {\bibfnamefont {G.~I.}\ \bibnamefont
  {Williams}},\ }\href {\doibase 10.1107/S0365110X55001527} {\bibfield
  {journal} {\bibinfo  {journal} {Acta. Cryst.}\ }\textbf {\bibinfo {volume}
  {8}},\ \bibinfo {pages} {494} (\bibinfo {year} {1955})}\BibitemShut {NoStop}%
\bibitem [{\citenamefont {Hill}(1970)}]{Hill70}%
  \BibitemOpen
  \bibfield  {author} {\bibinfo {author} {\bibfnamefont {H.~H.}\ \bibnamefont
  {Hill}},\ }\href@noop {} {\bibfield  {journal} {\bibinfo  {journal} {Nucl.
  Metall.}\ }\textbf {\bibinfo {volume} {17}},\ \bibinfo {pages} {2} (\bibinfo
  {year} {1970})}\BibitemShut {NoStop}%
\bibitem [{\citenamefont {Moore}\ and\ \citenamefont {{van der
  Laan}}(2009)}]{moorevanderlaan}%
  \BibitemOpen
  \bibfield  {author} {\bibinfo {author} {\bibfnamefont {K.~T.}\ \bibnamefont
  {Moore}}\ and\ \bibinfo {author} {\bibfnamefont {G.}~\bibnamefont {{van der
  Laan}}},\ }\href {\doibase 10.1103/RevModPhys.81.235} {\bibfield  {journal}
  {\bibinfo  {journal} {Rev. Mod. Phys.}\ }\textbf {\bibinfo {volume} {81}},\
  \bibinfo {pages} {235} (\bibinfo {year} {2009})}\BibitemShut {NoStop}%
\bibitem [{\citenamefont {McEwen}\ \emph {et~al.}(2007)\citenamefont {McEwen},
  \citenamefont {Walker}, \citenamefont {Le}, \citenamefont {McMorrow},
  \citenamefont {Colineau}, \citenamefont {Wastin}, \citenamefont {Wilkins},
  \citenamefont {Park}, \citenamefont {Bewley},\ and\ \citenamefont
  {Fort}}]{jmmm_upd3}%
  \BibitemOpen
  \bibfield  {author} {\bibinfo {author} {\bibfnamefont {K.~A.}\ \bibnamefont
  {McEwen}}, \bibinfo {author} {\bibfnamefont {H.~C.}\ \bibnamefont {Walker}},
  \bibinfo {author} {\bibfnamefont {M.~D.}\ \bibnamefont {Le}}, \bibinfo
  {author} {\bibfnamefont {D.~F.}\ \bibnamefont {McMorrow}}, \bibinfo {author}
  {\bibfnamefont {E.}~\bibnamefont {Colineau}}, \bibinfo {author}
  {\bibfnamefont {F.}~\bibnamefont {Wastin}}, \bibinfo {author} {\bibfnamefont
  {S.~B.}\ \bibnamefont {Wilkins}}, \bibinfo {author} {\bibfnamefont {J.-G.}\
  \bibnamefont {Park}}, \bibinfo {author} {\bibfnamefont {R.~I.}\ \bibnamefont
  {Bewley}}, \ and\ \bibinfo {author} {\bibfnamefont {D.}~\bibnamefont
  {Fort}},\ }\href {\doibase 10.1016/j.jmmm.2006.10.520} {\bibfield  {journal}
  {\bibinfo  {journal} {J. Mag. Mag. Mat.}\ }\textbf {\bibinfo {volume}
  {310}},\ \bibinfo {pages} {718} (\bibinfo {year} {2007})}\BibitemShut
  {NoStop}%
\bibitem [{\citenamefont {Martin-Martin}(2000)}]{amm00}%
  \BibitemOpen
  \bibfield  {author} {\bibinfo {author} {\bibfnamefont {A.}~\bibnamefont
  {Martin-Martin}},\ }\emph {\bibinfo {title} {Magnetism in Uranium
  Intermetallic Compounds}},\ \href@noop {} {Ph.D. thesis},\ \bibinfo  {school}
  {University College London} (\bibinfo {year} {2000})\BibitemShut {NoStop}%
\bibitem [{\citenamefont {Le}\ \emph {et~al.}(2012)\citenamefont {Le},
  \citenamefont {McEwen}, \citenamefont {Rotter}, \citenamefont {Jensen},
  \citenamefont {Bewley}, \citenamefont {Guidi},\ and\ \citenamefont
  {Fort}}]{upd3ins}%
  \BibitemOpen
  \bibfield  {author} {\bibinfo {author} {\bibfnamefont {M.~D.}\ \bibnamefont
  {Le}}, \bibinfo {author} {\bibfnamefont {K.~A.}\ \bibnamefont {McEwen}},
  \bibinfo {author} {\bibfnamefont {M.}~\bibnamefont {Rotter}}, \bibinfo
  {author} {\bibfnamefont {J.}~\bibnamefont {Jensen}}, \bibinfo {author}
  {\bibfnamefont {R.~I.}\ \bibnamefont {Bewley}}, \bibinfo {author}
  {\bibfnamefont {T.}~\bibnamefont {Guidi}}, \ and\ \bibinfo {author}
  {\bibfnamefont {D.}~\bibnamefont {Fort}},\ }\href {\doibase
  10.1088/0953-8984/24/3/036002} {\bibfield  {journal} {\bibinfo  {journal} {J.
  Phys.: Condens. Matter}\ }\textbf {\bibinfo {volume} {24}},\ \bibinfo {pages}
  {036002} (\bibinfo {year} {2012})}\BibitemShut {NoStop}%
\bibitem [{\citenamefont {Newman}\ and\ \citenamefont {Ng}(2000)}]{NN00}%
  \BibitemOpen
  \bibfield  {author} {\bibinfo {author} {\bibfnamefont {D.~J.}\ \bibnamefont
  {Newman}}\ and\ \bibinfo {author} {\bibfnamefont {B.~K.~C.}\ \bibnamefont
  {Ng}},\ }\href@noop {} {\emph {\bibinfo {title} {Crystal Field Handbook}}}\
  (\bibinfo  {publisher} {Cambridge University Press},\ \bibinfo {year}
  {2000})\BibitemShut {NoStop}%
\bibitem [{\citenamefont {McEwen}\ \emph {et~al.}(1998)\citenamefont {McEwen},
  \citenamefont {Steigenberger}, \citenamefont {Clausen}, \citenamefont
  {Kulda}, \citenamefont {Park},\ and\ \citenamefont {Walker}}]{MU+98}%
  \BibitemOpen
  \bibfield  {author} {\bibinfo {author} {\bibfnamefont {K.~A.}\ \bibnamefont
  {McEwen}}, \bibinfo {author} {\bibfnamefont {U.}~\bibnamefont
  {Steigenberger}}, \bibinfo {author} {\bibfnamefont {K.~N.}\ \bibnamefont
  {Clausen}}, \bibinfo {author} {\bibfnamefont {J.}~\bibnamefont {Kulda}},
  \bibinfo {author} {\bibfnamefont {J.-G.}\ \bibnamefont {Park}}, \ and\
  \bibinfo {author} {\bibfnamefont {M.~B.}\ \bibnamefont {Walker}},\ }\href
  {\doibase 10.1016/S0304-8853(97)00993-1} {\bibfield  {journal} {\bibinfo
  {journal} {J. Mag. Mag. Mat.}\ }\textbf {\bibinfo {volume} {177--181}},\
  \bibinfo {pages} {37} (\bibinfo {year} {1998})}\BibitemShut {NoStop}%
\bibitem [{\citenamefont {Gehring}(2007)}]{gilliannote}%
  \BibitemOpen
  \bibfield  {author} {\bibinfo {author} {\bibfnamefont {G.~A.}\ \bibnamefont
  {Gehring}},\ }\href@noop {} {} (\bibinfo {year} {2007}),\ \bibinfo {note}
  {private communication}\BibitemShut {NoStop}%
\bibitem [{\citenamefont {Rotter}\ \emph {et~al.}(1998)\citenamefont {Rotter},
  \citenamefont {M\"uller}, \citenamefont {Gratz}, \citenamefont {Doerr},\ and\
  \citenamefont {Loewenhaupt}}]{RM+98}%
  \BibitemOpen
  \bibfield  {author} {\bibinfo {author} {\bibfnamefont {M.}~\bibnamefont
  {Rotter}}, \bibinfo {author} {\bibfnamefont {H.}~\bibnamefont {M\"uller}},
  \bibinfo {author} {\bibfnamefont {E.}~\bibnamefont {Gratz}}, \bibinfo
  {author} {\bibfnamefont {M.}~\bibnamefont {Doerr}}, \ and\ \bibinfo {author}
  {\bibfnamefont {M.}~\bibnamefont {Loewenhaupt}},\ }\href {\doibase
  10.1063/1.1149009} {\bibfield  {journal} {\bibinfo  {journal} {Rev. Sci.
  Instr.}\ }\textbf {\bibinfo {volume} {69}},\ \bibinfo {pages} {2742}
  (\bibinfo {year} {1998})}\BibitemShut {NoStop}%
\bibitem [{\citenamefont {Zochowski}\ and\ \citenamefont
  {McEwen}(1994)}]{ZM94}%
  \BibitemOpen
  \bibfield  {author} {\bibinfo {author} {\bibfnamefont {S.~W.}\ \bibnamefont
  {Zochowski}}\ and\ \bibinfo {author} {\bibfnamefont {K.~A.}\ \bibnamefont
  {McEwen}},\ }\href {\doibase 10.1016/0921-4526(94)91855-4} {\bibfield
  {journal} {\bibinfo  {journal} {Physica B}\ }\textbf {\bibinfo {volume}
  {199\&200}},\ \bibinfo {pages} {416} (\bibinfo {year} {1994})}\BibitemShut
  {NoStop}%
\bibitem [{\citenamefont {Rotter}(2004)}]{rotter04}%
  \BibitemOpen
  \bibfield  {author} {\bibinfo {author} {\bibfnamefont {M.}~\bibnamefont
  {Rotter}},\ }\href {\doibase 10.1016/j.jmmm.2003.12.1394} {\bibfield
  {journal} {\bibinfo  {journal} {J. Magn. Mag. Mat.}\ }\textbf {\bibinfo
  {volume} {272--276}},\ \bibinfo {pages} {E481} (\bibinfo {year}
  {2004})}\BibitemShut {NoStop}%
\bibitem [{\citenamefont {Rotter}\ \emph {et~al.}()\citenamefont {Rotter} \emph
  {et~al.}}]{mcphasewebsite}%
  \BibitemOpen
  \bibfield  {author} {\bibinfo {author} {\bibfnamefont {M.}~\bibnamefont
  {Rotter}} \emph {et~al.},\ }\href@noop {} {}\bibinfo {note}
  {\url{http://www.mcphase.de}}\BibitemShut {NoStop}%
\bibitem [{\citenamefont {Cowley}\ and\ \citenamefont
  {Jensen}(1992)}]{cowleyjensener}%
  \BibitemOpen
  \bibfield  {author} {\bibinfo {author} {\bibfnamefont {R.~A.}\ \bibnamefont
  {Cowley}}\ and\ \bibinfo {author} {\bibfnamefont {J.}~\bibnamefont
  {Jensen}},\ }\href {\doibase 10.1088/0953-8984/4/48/021} {\bibfield
  {journal} {\bibinfo  {journal} {Journal of Physics: Condensed Matter}\
  }\textbf {\bibinfo {volume} {4}},\ \bibinfo {pages} {9673} (\bibinfo {year}
  {1992})}\BibitemShut {NoStop}%
\bibitem [{\citenamefont {Jensen}\ and\ \citenamefont
  {Mackintosh}(1991)}]{jensen}%
  \BibitemOpen
  \bibfield  {author} {\bibinfo {author} {\bibfnamefont {J.}~\bibnamefont
  {Jensen}}\ and\ \bibinfo {author} {\bibfnamefont {A.~R.}\ \bibnamefont
  {Mackintosh}},\ }\href@noop {} {\emph {\bibinfo {title} {Rare Earth
  Magnetism}}}\ (\bibinfo  {publisher} {Clarendon Press},\ \bibinfo {year}
  {1991})\BibitemShut {NoStop}%
\bibitem [{not()}]{note1}%
  \BibitemOpen
  \href@noop {} {}\bibinfo {note} {For a two level system, the mean field
  susceptibility is inversely proportional to
  $\Delta(\Delta-2n_{sd}|\bra{d}\hat{O}\ket{s}|^2\mathcal{J}(\mathbf{q}))$,
  which may be equated with the Curie-Weiss susceptibility, $\chi \propto
  1/(T-T_N)$. At low temperatures, $T\ll \Delta$, the thermal population factor
  $n_{sd}\approx\Delta/k_BT$, so that
  $T_N\appropto|\bra{d}\hat{O}\ket{s}|^2\mathcal{J}(\mathbf{q})$.}\BibitemShut
  {Stop}%
\bibitem [{\citenamefont {Kennedy}\ and\ \citenamefont {Eberhart}(1995)}]{pso}%
  \BibitemOpen
  \bibfield  {author} {\bibinfo {author} {\bibfnamefont {J.}~\bibnamefont
  {Kennedy}}\ and\ \bibinfo {author} {\bibfnamefont {R.}~\bibnamefont
  {Eberhart}},\ }in\ \href {\doibase 10.1109/ICNN.1995.488968} {\emph {\bibinfo
  {booktitle} {Proc. IEEE International Conference on Neural Networks,
  1995.}}},\ Vol.~\bibinfo {volume} {4}\ (\bibinfo {year} {1995})\ pp.\
  \bibinfo {pages} {1942--1948 vol.4}\BibitemShut {NoStop}%
\bibitem [{\citenamefont {McEwen}\ \emph {et~al.}(1993)\citenamefont {McEwen},
  \citenamefont {Steigenberger},\ and\ \citenamefont {Martinez}}]{MSM93}%
  \BibitemOpen
  \bibfield  {author} {\bibinfo {author} {\bibfnamefont {K.~A.}\ \bibnamefont
  {McEwen}}, \bibinfo {author} {\bibfnamefont {U.}~\bibnamefont
  {Steigenberger}}, \ and\ \bibinfo {author} {\bibfnamefont {J.~L.}\
  \bibnamefont {Martinez}},\ }\href {\doibase 10.1016/0921-4526(93)90669-W}
  {\bibfield  {journal} {\bibinfo  {journal} {Physica B}\ }\textbf {\bibinfo
  {volume} {186--188}},\ \bibinfo {pages} {670} (\bibinfo {year}
  {1993})}\BibitemShut {NoStop}%
\bibitem [{\citenamefont {Lingg}\ \emph {et~al.}(1999)\citenamefont {Lingg},
  \citenamefont {Maurer}, \citenamefont {M\"uller},\ and\ \citenamefont
  {McEwen}}]{LMMM99}%
  \BibitemOpen
  \bibfield  {author} {\bibinfo {author} {\bibfnamefont {N.}~\bibnamefont
  {Lingg}}, \bibinfo {author} {\bibfnamefont {D.}~\bibnamefont {Maurer}},
  \bibinfo {author} {\bibfnamefont {V.}~\bibnamefont {M\"uller}}, \ and\
  \bibinfo {author} {\bibfnamefont {K.~A.}\ \bibnamefont {McEwen}},\ }\href
  {\doibase 10.1103/PhysRevB.60.R8430} {\bibfield  {journal} {\bibinfo
  {journal} {Phys. Rev. B}\ }\textbf {\bibinfo {volume} {60}},\ \bibinfo
  {pages} {R8430} (\bibinfo {year} {1999})}\BibitemShut {NoStop}%
\bibitem [{\citenamefont {Fern\'andez-Rodr\'iguez}\ \emph
  {et~al.}(2010)\citenamefont {Fern\'andez-Rodr\'iguez}, \citenamefont
  {Lovesey},\ and\ \citenamefont {Blanco}}]{lovesey2010}%
  \BibitemOpen
  \bibfield  {author} {\bibinfo {author} {\bibfnamefont {J.}~\bibnamefont
  {Fern\'andez-Rodr\'iguez}}, \bibinfo {author} {\bibfnamefont {S.~W.}\
  \bibnamefont {Lovesey}}, \ and\ \bibinfo {author} {\bibfnamefont {J.~A.}\
  \bibnamefont {Blanco}},\ }\href {\doibase 10.1088/0953-8984/22/2/022202}
  {\bibfield  {journal} {\bibinfo  {journal} {J. Phys.: Condens. Matter}\
  }\textbf {\bibinfo {volume} {22}},\ \bibinfo {pages} {022202} (\bibinfo
  {year} {2010})}\BibitemShut {NoStop}%
\bibitem [{\citenamefont {Doerr}\ \emph {et~al.}(2005)\citenamefont {Doerr},
  \citenamefont {Rotter},\ and\ \citenamefont
  {Lindbaum}}]{rotterdoerrlindbaum}%
  \BibitemOpen
  \bibfield  {author} {\bibinfo {author} {\bibfnamefont {M.}~\bibnamefont
  {Doerr}}, \bibinfo {author} {\bibfnamefont {M.}~\bibnamefont {Rotter}}, \
  and\ \bibinfo {author} {\bibfnamefont {A.}~\bibnamefont {Lindbaum}},\ }\href
  {\doibase 10.1080/00018730500037264} {\bibfield  {journal} {\bibinfo
  {journal} {Advances in Physics}\ }\textbf {\bibinfo {volume} {54}},\ \bibinfo
  {pages} {1} (\bibinfo {year} {2005})}\BibitemShut {NoStop}%
\bibitem [{\citenamefont {Morin}\ and\ \citenamefont {Schmitt}(1990)}]{MS90}%
  \BibitemOpen
  \bibfield  {author} {\bibinfo {author} {\bibfnamefont {P.}~\bibnamefont
  {Morin}}\ and\ \bibinfo {author} {\bibfnamefont {D.}~\bibnamefont
  {Schmitt}},\ }in\ \href@noop {} {\emph {\bibinfo {booktitle} {Ferromagnetic
  Materials vol. 5}}},\ \bibinfo {editor} {edited by\ \bibinfo {editor}
  {\bibfnamefont {K.~H.~J.}\ \bibnamefont {Buschow}}\ and\ \bibinfo {editor}
  {\bibfnamefont {E.~P.}\ \bibnamefont {Wohlfarth}}}\ (\bibinfo  {publisher}
  {Elsevier},\ \bibinfo {address} {Amsterdam},\ \bibinfo {year} {1990})\ pp.\
  \bibinfo {pages} {1--132}\BibitemShut {NoStop}%
\end{thebibliography}

%


\end{document}